\documentclass{article}
\usepackage{amsmath}

\usepackage{arxiv}

\usepackage[utf8]{inputenc} 
\usepackage[T1]{fontenc}    
\usepackage{hyperref}       
\usepackage{url}            
\usepackage{booktabs}       
\usepackage{amsfonts}       
\usepackage{nicefrac}       
\usepackage{microtype}      
\usepackage{lipsum}		
\usepackage{graphicx}

\title{ Finslerian wormhole solution in the framework of modified gravity}

\author{Manjunath Malligawad$^1$ S. K. Narasimhamurthy$^{1*}$ Z. Nekouee$^2$ Mallikarjun Y Kumbar$^3$\\$^1$Department of PG Studies and Research in Mathematics,
	Kuvempu University, Jnana Sahyadri, \\Shankaraghatta - 577 451, Shivamogga, Karnataka, India.\\$^2$School of Physics, Damghan University, Damghan, 3671641167, Iran.\\$^3$Department of Mathematics,  Mahantswamy Arts, Science and Commerce College,\\  Haunsbhavi - 581 109, Haveri, Karnataka, India.\\
	\texttt{Corresponding author$^*$: nmurthysk@gmail.com} \\
}

\hypersetup{pdfkeywords={$f(R,T)$ gravity, Traversable Wormhole, Energy conditions, Exponential shape function, and Finsler Space-time.}
}

\begin{document}
\maketitle

\begin{abstract}
This article investigates the properties of a wormhole model in a specific gravity theory, namely $f(Ric, T)=Ric+2\lambda T$. The wormhole solution is analyzed using an exponential shape function. The study examines various parameters, such as density, radial pressure, transverse pressure, equation-of-state parameters, and energy conditions, within the framework of deformed gravity. The research emphasizes the influence of the parameter $\lambda$ on energy condition violations and the equilibrium state of the Finslerian wormhole solution. These effects are attributed to anisotropic and hydrostatic forces present in modified gravity. The study demonstrates that the gravity model effectively captures the characteristics of wormholes within the Finslerian space-time.
Additionally, the identified features of the wormhole are utilized to visualize its structure by creating a three-dimensional representation of the embedded surface. In summary, this research contributes to understanding wormholes in modified gravity theories, highlighting the importance of the parameter $\lambda$ in determining their behavior and properties.
\end{abstract}


\section{Introduction}\label{sec.1}

The concept of wormholes was initially introduced by the German mathematician  Weyl H \cite{Weyl}. Weyl described a wormhole as an asymptotically flat tube-like structure, characterizing it as a tunnel with two entrances or connecting two distinct points in different space-times, essentially forming a bridge that joins two ends of the same space-time. Consequently, the topic of wormholes captured the interest of numerous geometricians and physicists. The Schwarzschild wormhole, pioneered by Schwarzschild, represents the first type of wormhole solution in this field of study.\\

{Wormholes stand out as the most widely recognized and extensively explored concepts in both General Relativity (GR) and alternative theories of gravity. Solutions derived from GR give rise to these theoretical shortcuts, commonly referred to as Einstein-Rosen bridges, which were initially mathematically postulated by A. Einstein et al. \cite{Einstein}. However, subsequent research by R. W. Fuller et al. \cite{Fuller}, published in 1962, demonstrated the instability of this type of wormhole. Specifically, if it connects two regions within the same universe, any object moving slower than the speed of light that traverses from one exterior area to the other is susceptible to falling out. Wormholes can be broadly categorized into two types: static and dynamic wormholes.} \\

{The concept of traversable wormholes, as proposed by M. S. Morris et al. \cite{Morris}, explores the idea of using wormholes for time travel or interstellar travel, distinct from the Einstein-Rosen bridge theory. To maintain wormhole geometry, exotic matter is crucial in the throat, exerting outward pressure to prevent collapse and ensuring stability for traversal. Researchers focus on minimizing violations of the null energy condition to enhance stability. D. Hochberg et al. \cite{Hochberg} showed that all types of traversable wormholes violate energy conditions, highlighting the challenge of addressing this violation while studying their rigidity.
	This is vital for advancing our understanding of these hypothetical structures.} \\

The energy conditions connected are DEC $\subset$ WEC $\subset$ NEC and SEC $\subset$ NEC. Violating the NEC involves violating all of the energy conditions. In reality, even under conditions weaker than the NEC \cite{Galloway}, traversable wormholes cannot be realized in GR according to the following topological censorship theorem in $f(R)$ gravity \cite{Friedman}, traversable wormholes were examined by  F. S. N. Lobo et al. \cite{Lobo}. They supported the wormhole constructions and identified the causes of the null energy condition's dissatisfaction. $f(R)$ and $f(R, T)$ theories have received a lot of attention recently as a way to explain several elements of the universe. Errehymy et al. \cite{Errehymy} delved into the world of generalized Ellis-Bronikov (GEB) traversable wormholes specifically, in the context of static and spherically symmetric space-time governed by $f(R)$ gravity.
Shifting the cosmic focus, spherically symmetric traversable wormholes took the spotlight in discussions on torsion-matter coupling gravity \cite{Errehymy1}. By adding the stress-energy tensor term,  T. Harko et al. \cite{Harko} generalized the $f(R)$ gravitational theory and developed the $f(R, T)$ MGT. Gravity models depend on the source term since matter and gravity are connected. Using $f(R, T)$ gravity,  E. H. Baffou et al.  \cite{Baffou} investigated the cosmic evolution of deceleration and the equation of state (EoS) parameters. In the context of the FLRW model, in the study of static wormholes in the $f(R, T)$ theory of gravity, P. H. R. S. Moraes et al. \cite{Sahoo} provided some wormhole models with various matter content assumptions.   {S. Najafi et al. \cite{Najafi} investigated traversable wormholes with one extra space-like compact dimension. In this respect, they considered how it affected energy density, scale factor, and shape function. João Luís Rosa et al. \cite{Paul1, Nailya} examined traversable wormhole solutions within the linear context $f(R, T)$ gravity.}

Mustafa et al. \cite{Mustafa} navigated through the landscape of wormhole geometry using conformal symmetry within the framework of $f(\mathcal{T}, T)$ gravity.
As if that wasn't intriguing enough, reference \cite{Mustafa1} ventured into exploring wormhole solutions in the galactic halo region, this time under the influence of symmetric teleparallel gravity.
{In $f(R, T)$ modified gravity theory (MGT)~\cite{Moraes1}, describe the geometry of the wormhole by using an exponential shape function. The exponential $f(R, T)$ MGT solution is provided by P. H. R. S. Moraes et al. \cite{Moraes2}, and the concept of exponential $f(R, T)$ MGT is considered original in literature.
	In recent studies on $f(R, T)$ modified gravity, various researchers have explored different concepts to derive astrophysical solutions.
	Ayan Banerjee \cite{Banerjee}, for instance, utilized the spherical symmetry of wormholes and the presence of a conformal Killing symmetry to obtain more refined $f(R, T)$ solutions.
	R. Mudassar et al. \cite{Mudassar} delved into the thermodynamics of spherically symmetric and static traversable wormholes, encompassing scenarios such as Morris–Thorne wormholes and charged wormholes within the framework of $f(R, T)$ gravity.
	N. Godani \cite{Godani}, on the other hand, investigated traversable wormholes in the context of $f(R, T)$ theory with a specific functional form, $f(R, T) = R + 2\lambda T$.
	M. Z. Bhatti et al. \cite{Bhatti} explored thin-shell wormholes within the $f(R, T)$ gravity framework. These studies collectively contribute to a comprehensive understanding of the implications and applications of $f(R, T)$ modified gravity in the context of wormhole physics}.\\

{Recently, Finslerian geometry has caught the attention of many physicists because it explains a variety of issues that Einstein's gravity is unable to explain. Riemannian geometry is one of the particular cases of Finsler geometry. Modern standard high-energy theories and Finsler-like gravity theories are consistent with experimental results.
	Without considering the dark matter hypothesis, Finsler geometry provides a better method for addressing problems with experimental results of flat rotation curves and spiral galaxies \cite{Chang}.
	Nicoleta Voicu et al. \cite{Voicu} have summarized recent advancements in using Finsler geometry as the space-time geometric framework extension.
	Claus Lämmerzahl et al. \cite{Claus} have delved into Finslerian adaptations of Maxwell's equations, the Klein-Gordon, and the Dirac equations.
	Additionally, Lämmerzahl has investigated experimental tests to validate Finslerian gravity theories.
	Vacaru et al. \cite{Sergiu} work has focused on establishing the geometric and physical foundations of Finsler gravity theories, particularly those employing metric-compatible connections on tangent bundles or (pseudo) Riemannian manifolds endowed with nonholonomic frame structures. Abdel Nasser Tawfik's contributions lie in the conceptualization of Born reciprocity within discretized Finsler structures. Tawfik et al. \cite{Tawfik} have explored approaches to quantizing curvature tensors in the context of GR on three-sphere models. Furthermore, Tawfik et al. \cite{Tawfik1} have investigated the implications of Born reciprocity and the relativistic generalized uncertainty principle within Finsler structures, analyzing their influence on emerged curvatures and their role in regulating space-time and addressing initial singularities. Additionally, Tawfik et al. \cite{Tawfik2} have examined the behavior of timelike geodesic congruence within the simplest solutions derived from the Einstein field equations.
	B. Sumita et al. \cite{Sumita} introduced the concept of a gravastar within the context of Finslerian space-time, offering an alternative notion to the conventional Finslerian black hole.
	D. P. Krishna et al. \cite{Krishna} investigated the potential existence of a traversable wormhole utilizing the framework of Finsler–Randers (F–R) geometry.
	Additionally, J. Praveen et al. \cite{Praveen} extensively examined the phenomenon of cosmological constant roll inflation, analyzing it within the framework of Finsler-Barthel-Kropina geometry.
	These studies collectively deepen our understanding of Finsler geometry's applicability in gravitational theories, ranging from theoretical developments to experimental verifications and implications for fundamental principles like curvature quantization and space-time singularities.}\\

In Finslerian geometry, the traversable wormhole was created by F.  Rahman et al. \cite{Rahman}.
They discovered an exact solution for various types of shape functions, red-shift functions, and EoS.
They also talked about the features of wormhole models.
Studying whether or if wormhole stable solutions exist without the need for exotic matter, wormhole geometry has also been investigated thoroughly using modified theories, where MGTs allow us to discuss the issue of exotic matter \cite{Clifton, Yashwanth}.\\

{In the realm of Finsler geometry, H. M. Manjunatha and colleagues, as outlined in their work \cite{Manjunatha}, delve into the exploration of wormhole models within Einstein Finsler gravity, specifically with an exponential shape function. Their investigation extends to analyzing the gravitational field equation to derive wormhole solutions in the Finslerian framework, incorporating considerations for an anisotropic energy-momentum tensor. Expanding the scope within Finsler geometry, researchers like Nekouee et al. \cite{Nekouee1} are unraveling cosmic mysteries using the Finsler-Randers (FR) metric. Additionally, attention has been directed towards the study of Finslerian Schwarzschild-de Sitter space-time, as highlighted in the research by Manjunatha et al. \cite{Manjunatha1}. This exploration resembles a process of peeling back the layers of the universe through the lens of geometry. Inspired by the exponential $f(R)$ gravity model, researchers have resolved wormhole field equations using the Finslerian $f(R, T)$ Modified Gravity Theory (MGT) formalism. This approach is particularly applied to the family of $f(R,T)$ MGT, with the specific form $f(R,T) = R+f(T)$, where $f(T)=2\lambda T$, with $R$ representing the Ricci scalar, $T$ representing the energy-momentum tensor, and $\lambda$ being a parameter.}
In this article, we explore the Finslerian wormhole model under $f(R, T)$ MGT with exponential shape function and examine the wormhole solutions based on parameter values that indicate the violation of energy conditions.\\

The following planning guides the structure of the paper. In section (\ref{sec2}), we quickly review the ideology of Finsler geometry.  In section (\ref{sec3}), we described the Finsler geometry formalization in $f(Ric,T)$ MGT and discussed the energy conditions. In section (\ref{sec4}), we discussed and analyzed the obtained outcomes. The study concludes with section (\ref{sec5}), and we provide some concluding observations.

\section{Preliminaries and Notations of Finsler Geometry}\label{sec2}
\par Finsler structure on manifold $M$ is defined as function $F: TM\rightarrow [0, \infty)$  which satisfies the below properties:\\
1. Regularity: $F$ is smooth function on the $TM\backslash\{0\}$.\\
2. Positive Homogeneity: $F(x, cy)= cF(x, y)$ for all $c>0$.\\
3. Strong Convexity: The $n\times n$ Hessian matrix
\begin{equation}\label{eq.1}
	g_{\mu\nu}=\frac{\partial^{2}\left(\frac{1}{2}F^2\right)}{\partial y^{\mu}\partial y^{\nu}}=\frac{1}{2}\dot{\partial}_{\mu}\dot{\partial}_{\nu}F^{2},
\end{equation}
Finsler structure $F$ is positive definite on $TM\backslash\{0\}$, and it is the function of ($x^{i}, y^{i}$). The pair ($M, F$) is called Finsler space.

A Finslerian metric is referred to as Riemannian if $F^{2}$ is quadratic in $y$. For the Finsler manifold, the geodesic equation is as follows:
\begin{equation}\label{eq.2}
	\frac{d^{2}x^{\mu}}{d\tau^{2}}+2G^{\mu}=0,
\end{equation}
where
\begin{equation}\label{eq.3}
	G^{\mu}=\frac{1}{4}g^{\mu\nu}\left(\frac{\partial^{2}F^{2}}{\partial x^{\upsilon}\partial y^{\nu}}y^{\upsilon}-\frac{\partial F^{2}}{\partial x^{\nu}}\right),
\end{equation}
are called geodesic spray coefficients. \\

We observed that the Finslerian structure $F(x,y)$ along the geodesic is constant. The Finslerian modified Ricci tensor equation suggested by Akbar-Zadeh~\cite{Akbar-Zadeh} is as follows.
\begin{equation}\label{eq.4}
	Ric_{\mu\nu}=\frac{\partial^{2}\left(\frac{1}{2}F^2 Ric\right)}{\partial y^{\mu}\partial y^{\nu}},
\end{equation}
here, $Ric$ stands for Ricci scalar. $Ric$ is a geometric invariant with the following expression
\begin{equation}\label{eq.5}
	Ric=g^{\mu\nu}R_{\mu\nu}.
\end{equation}
on any basis, equation (\ref{eq.5}) is true. Moreover, in equation (\ref{eq.5}), $R_{\mu\nu}$ is a representation of the flag's original curvature. It might be stated as,
\begingroup\makeatletter\def\f@size{10}\check@mathfonts
\begin{equation}\label{eq.6}
	R^\mu_\nu=\frac{1}{F^2}\left(2\frac{\partial G^\mu}{\partial x^\nu}-y^\upsilon \frac{\partial^2 G^\mu}{\partial x^\upsilon \partial y^\nu}+2G^\upsilon \frac{\partial^2 G^\mu}{\partial y^\upsilon \partial y^\nu}-\frac{\partial G^\mu}{\partial y^\upsilon} \frac{\partial G^\upsilon}{\partial y^\nu} \right).
\end{equation}
\endgroup
Thus, $Ric$ is as follows,
\begingroup\makeatletter\def\f@size{11}\check@mathfonts
{\begin{eqnarray}\label{eq.7}
		Ric&\equiv R^\mu_\mu
		= \frac{1}{F^2}\left(2\frac{\partial G^\mu}{\partial x^\mu}-y^\upsilon \frac{\partial^2 G^\mu}{\partial x^\upsilon \partial y^\mu}+2G^\upsilon \frac{\partial^2 G^\mu}{\partial y^\upsilon \partial y^\mu}- \frac{\partial G^\mu}{\partial y^\upsilon} \frac{\partial G^\upsilon}{\partial y^\mu}\right).
\end{eqnarray}}
\endgroup
Finslerian modified formula for scalar curvature is
\begin{equation}\label{eq.8}
	S=g^{\mu\nu}Ric_{\mu\nu},\hspace{0.7cm}
\end{equation}
and Einstein tensor formula is
\begin{equation}\label{eq.9}
	G_{\mu\nu}=Ric_{\mu\nu}-\frac{1}{2}g_{\mu\nu}S.
\end{equation}
Since it is derived from $Ric$, connections do not affect the Finslerian-modified Einstein tensor. It is solely reliant on the Finslerian structure. As a result, the Finslerian gravitational field equations are also insensitive to the connections.\\

{Birkhoff theorem \cite{Morris, Li1, Chowdhury} shows well that most static vacum solutions are reducible to Schwarzschild form. Hence, we will assume the Finsler structure in a similar form,}
\begin{equation}\label{eq.10}
	F^2=e^{2a(r)}y^t y^t-\left (1-\frac{b(r)}{r}\right)^{-1}y^r y^r-r^2\bar{F}^{2}(\theta, \phi, y^\theta, y^\phi),
\end{equation}
where $\bar{F}$ is a Finsler structure in two dimensions \cite{Li1, Wang}, and we assume that $\bar{F}$ has the form,
\begin{equation}\label{eq.11}
	\bar{F}^2=y^\theta y^\theta+A(\theta,\phi)y^\phi y^\phi.
\end{equation}
{Within the Finsler framework in equation (\ref{eq.10}),  $a(r)$ stands for the redshift function, while $b(r)$ is the shape function \cite{Konoplya}. $a(r)$ explains the redshift effect and tidal force in the wormhole space-time. Specifically, $a(r)$ provides information about how time dilation and gravitational effects vary with radial distance in the wormhole.
	$a(r)$ must be finite everywhere to avoid the presence of horizons within the wormhole. The finiteness of $a(r)$ is essential to ensure that the wormhole remains traversable and free from singularities or event horizons. Since a horizon, if present, would prevent two-way travel through the wormhole. Particularly when $a(r)=0$ \cite{Mishra,Manjunatha}, implies a tideless condition. In the context of wormholes, this tideless condition is a significant characteristic, as it indicates that there are no significant tidal forces affecting the space-time \cite{Konoplya,Krishna}.}
The shape function  $b(r)$ specifies how a wormhole is shaped. The wormhole throat is positioned at a minimal surface radius $b(r_0) = r_0$, and the nonmonotonic radial coordinate $r$ falls from infinity to $r_0$ before increasing from $r_0$ back to infinity.  $b(r)$ needs to meet the requirements Morris-Thorne et al. \cite{Morris} below to produce the wormhole solutions:
\begin{itemize}
	\item Radial co-ordinate $r$ has a range of values from $r_0\leq r \leq \infty $, where $r_0$ is the  throat radius.
	\item Within the throat $b(r)$ fulfill the requirement  $b(r_0)= r_0$, as well as for out of the throat, i.e., for $r > r_0, 0<1-\frac{b(r)}{r}$.
	\item Shape function $b(r)$ at the throat must satisfy the flaring out requirement, i.e., $b^{\prime}(r_0) < 1$,
	where the derivative is with respect to $r$.
	\item The minimum requirement for the space-time geometry to be asymptotically flat $\frac{b(r)}{r}\rightarrow 0$  as $|r|\rightarrow \infty$.
\end{itemize}
The geometry of the wormhole space-time is described by the requirements listed above.\\

For two-dimensional Finsler structure $\bar{F}$ equation (\ref{eq.11}) one can obtain the Finslerian metric as,
\begin{equation}\label{eq.12}
	\bar{{g}}_{ij}=diag(1,A(\theta,\phi)),\hspace{0.5cm}
\end{equation}
\begin{equation}\label{eq.13}
	\bar{{g}}^{ij}=diag\left(1,\frac{1}{A(\theta,\phi)}\right),
\end{equation}
{where $\bar{g}_{ij}$ and $\bar{g}^{ij}$  are the metrics derived from $\bar{F}$ and the indices $i, j$ run over the angular coordinates $\theta, \phi$.} The $\bar{R}ic$ of the Finslerian structure $\bar{F}$ can be found by applying the formula from the reference F. Rahman et al. \cite{Rahman}.
\begin{equation}\label{eq.14}
	\bar{R}ic=\frac{1}{2A}\left[-\frac{\partial^2A}{\partial\theta^2}+\frac{1}{2A}\left(\frac{\partial A}{\partial \theta}\right)^2\right].
\end{equation}
For the Finsler structure $\bar{F}$, we can better understand wormholes in the Finslerian theory by using the  $\bar{R}ic=\eta$. Here, it is assumed to be constant. In three separate circumstances, we find the solution to the differential equation (\ref{eq.14}) by solving it $(\eta>0, ~ \eta=0, ~ \eta<0)$. Consequently, the Finsler structure $\bar{F}$ equation (\ref{eq.11}) changes to,
\begin{eqnarray}
\bar{F}^2=y^\theta y^\theta +C \sin^2(\sqrt{\eta}\theta)y^\phi y^\phi ~~~~~~~~~~~~~~  \textrm{for} ~~ (\eta>0),\\\label{eq.15}
\bar{F}^2=y^\theta y^\theta +C \theta^2 y^\phi y^\phi  \hspace{2.5cm}  \textrm{for} ~~ (\eta=0),\\\label{eq.16}
\bar{F}^2=y^\theta y^\theta +C \sinh^2(\sqrt{-\eta}\theta)y^\phi y^\phi ~~~~~~~~~~ \textrm{for} ~~ (\eta<0).\label{eq.17}
\end{eqnarray}
Now let's use $C=1$. For the solution $\eta>0$ the Finslerian wormhole structure equation (\ref{eq.10}) can be written as
{\begin{eqnarray}\label{eq.18}
		F^2=e^{2a(r)} y^t y^t-\left(1-\frac{b(r)}{r}\right)^{-1}y^ry^r-
		r^2\bigg(y^\theta y^\theta+\sin^2(\sqrt{\eta}\theta)y^\phi y^\phi\bigg).
	\end{eqnarray}
	There are two distinct classifications of Finsler spaces. One classification involves Riemannian spaces, where a Finslerian metric is classified as Riemannian when $F^2$ exhibits a quadratic form in y. The second type is Randers space-time \cite{Randers}. The Riemannian metric is denoted by $\alpha$, and the appropriate Riemannian wormhole structure is provided by}
\begin{eqnarray}\label{eq.19}
	\alpha^2=e^{2a(r)}y^t y^t-\left(1-\frac{b(r)}{r}\right)^{-1} y^r y^r-
	r^2\bigg(y^\theta y^\theta+\sin^2(\theta)y^\phi y^\phi\bigg).
\end{eqnarray}
Using equation (\ref{eq.19}), we can written equation (\ref{eq.18}) as,
\begin{equation}\label{eq.20}
	F^2=\alpha^2+r^2(\sin^2\theta -\sin^2\sqrt{\eta}\theta)y^\phi y^\phi.
\end{equation}
If we take $r^2 (\sin^2\theta- \sin^2\sqrt{\eta}\theta) y^\phi y^\phi=\beta^2$ into consideration, then equation (\ref{eq.20}) yields,
\begin{equation}\label{eq.21}
	{F}^2=\alpha^2(1+s^2),
\end{equation}
where
\begin{equation*}
	s=\frac{\beta}{\alpha}=\frac{b_\phi y^\phi}{\alpha}.
\end{equation*}
We consider $b_\phi=r\sqrt{\sin^2\theta- \sin^2\sqrt{\eta}\theta}$. Additionally, the $\beta=b_\mu y^\mu$ denotes the differential 1-form where $b_\mu  = (0,0,0,b_\phi)$. Thus, equation (\ref{eq.21}) yields,
\begin{equation}\label{eq.22}
	F=\alpha\varphi(s),
\end{equation}
where $\varphi(s)=\sqrt{1+s^2}$. The Finslerian wormhole structure $F$ is indicated by equation (\ref{eq.22}), which represents the Finsler space with an $(\alpha,\beta)$-metric.\\

Concerning the Riemannian metric $\alpha$, the given symbol ``$|$" signifies the covariant derivative. The equations for $K_V(\alpha)$ and $K_V(\beta)$ are then provided by,
\begin{eqnarray*}
	K_V(\alpha)&=\frac{1}{2\alpha}(V_{\mu|\nu}+V_{\nu|\mu})y^\mu y^\nu,\\
	K_V(\beta)&=\left(V^\mu \frac{\partial b_\nu}{\partial x^\mu}+ b_\mu\frac{\partial V^\mu}{\partial x^\nu}\right) y^\nu.
\end{eqnarray*}
The Finsler space's Killing equation, $K_V (F)=0$, is derived from isometric transformations \cite{Li1}.
\begin{equation}\label{eq.23}
	\left(\varphi(s)-s\frac{\partial \varphi (s)}{\partial s}\right)K_V(\alpha)+\frac{\partial \varphi (s)}{\partial s} K_V(\beta)=0.
\end{equation}
From equation (\ref{eq.22}), the Killing equation (\ref{eq.23}) becomes,
\begin{equation*}
	\alpha K_V(\alpha)+\beta K_V(\beta)=0,
\end{equation*}
and it pays off
\begin{equation}\label{eq.24}
	K_V(\alpha)=0 ~~ \textrm{and} ~~ K_V(\beta)=0
\end{equation}
as a result, we have
\begin{equation}\label{eq.25}
	V_{\mu|\nu}+V_{\nu|\mu}=0,
\end{equation}
\begin{equation}\label{eq.26}
	V^\mu\frac{\partial b_\nu}{\partial x^\mu}+b_\mu \frac{\partial V^\mu}{\partial x^\nu}=0.
\end{equation}
In Randers-Finsler space-time equation (\ref{eq.19}), the Killing equation (\ref{eq.25}) appears identical to the Riemannian counterpart.  {However, the additional Killing equation (\ref{eq.26}) acts as a constraint on equation (\ref{eq.25}). Consequently, the number of independent Killing vectors in Randers-Finsler space-time (\ref{eq.19}) is less compared to that in Riemannian space-time $\alpha$ \cite{XLi, Wang}. The Lie derivatives $L_V(\alpha) = 0$ and $L_V(\beta) = 0$, are comparable to the Killing  equations equation (\ref{eq.25}), and equation (\ref{eq.26}) respectively}. In addition, we have $\alpha^2=h_{\mu\nu}$. We can get the Riemannian wormhole metric as,
\begin{equation*}
	h_{\mu\nu}=diag\left(e^{2a(r)}, -\left(1-\frac{b(r)}{r}\right)^{-1}, -r^2, -r^2\sin^2 \theta\right).
\end{equation*}
The Riemannian space-time isometric symmetry is broken as a result of the Killing equation (\ref{eq.26}).\\

{The redshift function should be positive for every $r>r_0$ if there are no horizons or singularities. In our current study, the redshift function $a(r)$ is taken to be constant as a result, $a^{\prime}(r)=0$. One of the crucial properties of traversable wormholes is that the tidal gravitational forces that travelers experience must be small or $a^{\prime}(r)=0$. The zero-tidal-force solutions have been examined by M. S. Morris et al. \cite{Morris}, who also took into account the vanishing redshift function.} As a result, the Finslerian wormhole structure equation (\ref{eq.10}) becomes,
\begin{eqnarray}\label{eq.27}
	F^2= y^t y^t-\left(1-\frac{b(r)}{r}\right)^{-1}y^ry^r-
	r^2\bigg(y^\theta y^\theta+\sin^2(\sqrt{\eta}\theta)y^\phi y^\phi\bigg).
\end{eqnarray}

{As we know, the Schwarzschild metric is known to be the exterior metric for any compact object in the universe. From Birkhoff’s theorem, it is well-known that most static vacuum solutions are reducible to Schwarzschild’s form. Our obtained Finslerian metric is consistent with this exterior metric and matches the continuity conditions, which can be well established from the reference \cite{Li1, Chowdhury}.}
One can derive the Finsler metric as,
\begin{equation}\label{eq.28}
	g_{\mu\nu}=diag\left(1,-\left(1-\frac{b(r)}{r}\right)^{-1},-r^2, -r^2\sin^2(\sqrt{\eta}\theta)\right),
\end{equation}
\begin{equation}\label{eq.29}
	g^{\mu\nu}=diag\left(1,-\bigg(1-\frac{b(r)}{r}\bigg),\frac{-1}{r^2}, \frac{-1}{r^2\sin^2(\sqrt{\eta}\theta)}\right). \hspace{.8cm}
\end{equation}
{When looking at the Finslerian wormhole crucial role is played by the Ricci scalar $Ric=\eta$ of the two-dimensional Finsler structure $\bar{F}$. We conclude that the structure $F$ has a constant flag curvature because $\eta$=constant. The geodesic spray coefficients can be found by putting the Finslerian wormhole structure equation (\ref{eq.27}) in equation (\ref{eq.2}).}
\begin{equation}\label{eq.30}
	G^t=0, \hspace{7.5cm}
\end{equation}
\begin{equation}\label{eq.31}
	G^r=\frac{rb^{\prime}-b}{4r(r-b)}y^ry^r-\frac{r-b}{2} y^\theta y^\theta-\frac{r-b}{2}\sin^2(\sqrt{\eta}\theta)y^\phi y^\phi,
\end{equation}
\begin{equation}\label{eq.32}
	G^\theta=\frac{1}{r}y^ry^\theta-\frac{\sqrt{\eta}}{2}\sin(\sqrt{\eta}\theta)\cos(\sqrt{\eta}\theta)y^\phi y^\phi, \hspace{2.3cm}
\end{equation}
\begin{equation}\label{eq.33}
	G^\phi=\frac{1}{r}y^ry^\phi+\sqrt{\eta}\cot(\sqrt{\eta}\theta)y^\theta y^\phi. \hspace{3.8cm}
\end{equation}
By substituting the expressions from equations (\ref{eq.30}-\ref{eq.33}) in equation (\ref{eq.7}), we obtain,
\begin{eqnarray}\label{eq.34}
	F^2Ric=\frac{rb^{\prime}-b}{r^2(r-b)}y^ry^r+\left(\eta-1+\frac{b}{2r}+\frac{b^{\prime}}{2}\right) y^\theta y^\theta+ \nonumber \\ ~~~~~~~~~~\left(\eta-1+\frac{b}{2r}+\frac{b^{\prime}}{2}\right)\sin^2(\sqrt{\eta}\theta)y^\phi y^\phi.
\end{eqnarray}
From equation (\ref{eq.4}), we deduced Finslerian modified Ricci tensor components  as,
\begin{equation}\label{eq.35}
	Ric_{tt}=0,
\end{equation}
\begin{equation}\label{eq.36}
	Ric_{rr}=\frac{rb^{\prime}-b}{r^2(r-b)},
\end{equation}
\begin{equation}\label{eq.37}
	Ric_{\theta\theta}=\left(\eta-1+\frac{b}{2r}+\frac{b^{\prime}}{2}\right),
\end{equation}
\begin{equation}\label{eq.38}
	Ric_{\phi\phi}=\left(\eta-1+\frac{b}{2r}+\frac{b^{\prime}}{2}\right)\sin^2(\sqrt{\eta}\theta).
\end{equation}
By substituting expressions equation (\ref{eq.35}-\ref{eq.38}) and equation (\ref{eq.29}) in equation (\ref{eq.8}), we can determine the scalar curvature for the Finslerian wormhole space-time as
\begin{equation}\label{eq.39}
	S=-\frac{2}{r^2}(b^{\prime}+\eta-1).
\end{equation}
Using the components of the Finslerian-modified Ricci tensor and the scalar curvature from equation (\ref{eq.9}), one can quickly determine the components of the Finslerian-modified Einstein tensor as
\begin{equation}\label{eq.40}
	G^t_t=\frac{1}{r^2}(b^{\prime}+\eta-1),
\end{equation}
\begin{equation}\label{eq.41}
	G^r_r=\frac{b}{r^3}+\frac{1}{r^2}(\eta-1),
\end{equation}
\begin{equation}\label{eq.42}
	G^\theta_\theta=G^\phi_\phi=\frac{rb^{\prime}-b}{r^3}.\hspace{.3cm}
\end{equation}
From components of the Finslerian modified Einstein tensor equation (\ref{eq.40}-\ref{eq.42}) depend only on $r$ and are direction-independent. With the help of the formula, the Chern connection is determined by
\begin{equation}\label{eq.43}
	\Gamma^\alpha_{\nu\upsilon}=\frac{\partial^2G^\alpha}{\partial y^\nu \partial y^\upsilon}-A^\alpha_{\nu\upsilon| \beta}\frac{y^\beta}{F}.
\end{equation}
The term $A_{\mu\nu\upsilon}=g_{\mu\alpha} A^\alpha_{\nu\upsilon}$ in the above equation, which represents the Cartan connection, is defined as,
\begin{equation}\label{eq.44}
	A_{\mu\nu\upsilon}=\frac{F}{4}\frac{\partial^3F^2}{\partial y^\mu \partial y^\nu \partial y^\upsilon},
\end{equation}
for all indices $\mu,\nu$ and $\upsilon $, we obtain $A_{\mu\nu\upsilon}=0$, to the Finslerian wormhole structure equation (\ref{eq.27}). From this, we obtain that the Cartan connections for the Finslerian wormhole space-time vanish. As a result, we can calculate the Chern connection coefficients using equation (\ref{eq.43}). So we have,
\begin{eqnarray}\label{eq.45}
	\Gamma^t_{\mu\nu}&=&0 ~~~~~ (\textrm{for all indices} ~ \mu,\nu), ~~~ \Gamma^\theta_{r\theta}=\Gamma^\phi_{r\phi}=\frac{1}{r}, \nonumber\\
	\Gamma^r_{rr}&=&\frac{rb^{\prime}-b}{2r(r-b)}, ~~~~~~~~~~~~~~~ \Gamma^r_{rt}=\Gamma^r_{r\theta}=\Gamma^r_{r\phi}=0,\nonumber\\
	\Gamma^r_{\theta\theta}&=&-(r-b),~~~~~~~~~ \Gamma^r_{\phi\phi}=-(r-b)\sin^2(\sqrt{\eta}\theta),\nonumber\\
	\Gamma^\theta_{\theta\theta}&=&\Gamma^\theta_{\theta t}=\Gamma^\theta_{\theta\phi}=0,~~~~~~~~~~~~~~~~~~~~~~~~~~~~~~~~~~~~~\nonumber \\
	\Gamma^\phi_{\phi t}&=&\Gamma^\phi_{\phi\phi}=0, ~~~~~~~~~~~~~~~~~~\Gamma^\phi_{\phi\theta}=\sqrt{\eta}\cot(\sqrt{\eta}\theta).\nonumber\\
\end{eqnarray}
Now we define the derivative $\frac{\delta}{\delta x^\mu}$ as,
\begin{equation*}
	\frac{\delta}{\delta x^\mu}=\frac{\partial}{\partial x^\mu}-\frac{\partial G^\alpha}{\partial y^\mu}\frac{\partial}{\partial y^\alpha }.
\end{equation*}
We can determine the covariant derivative of $G^\mu_\nu$ by the following equation
\begin{equation}\label{eq.46}
	G^\mu_{\nu|\mu}= \frac{\delta G^\mu_\nu}{\delta x^\mu}+\Gamma^\mu_{\mu\alpha}G^\alpha_\nu-\Gamma^\alpha_{\mu\nu}G^\mu_\alpha.
\end{equation}
From equation (\ref{eq.46}) all the covariant derivative components becomes $G^\mu_{t|\mu}=G^\mu_{r|\mu}=G^\mu_{\theta|\mu}=G^\mu_{\phi|\mu}=0$.
This leads us to the conclusion that the covariant derivative of $G^\mu_\nu$ for the Finslerian modified Einstein tensor is conserved.\\
\section{Formalism of Finsler Geometry in $f(Ric,T)$ Gravity}\label{sec3}
The $f(R,T)$ MGT has received much attention in recent research and seems suitable for handling wormhole construction problems for understanding the observed universe. $f(R,T)$ MGT is the combination of two functions $R$ and $T$, where $R$ stands for the Ricci scalar and $T$ for the trace of the energy-momentum tensor, is frequently studied in the research T. Harko \cite{Harko}. Since the distribution of the matter needed to create wormholes remains a problem for physicists.  {So, we take the  universal anisotropic energy-momentum tensor~\cite{Mak},
	\begin{equation}\label{eq.47}
		T^\mu_\nu=(\rho+p_t)u^\mu u_\nu+(p_r-p_t)x^\mu x_\nu-p_t \delta^\mu_\nu.
\end{equation}}
When taken in the radial direction, energy density is represented by $\rho$, $u^\mu$ is the four-velocity such that $u^\mu u_\mu=1$ and $x^\mu$ is the space-like unit vector, such that $x^\mu x_\mu=-1$. $p_r$ is the radial pressure, determined by the direction of the space-like unit vector $x^{\mu}$, where $p_t$ represents the transverse pressure measures perpendicular to $x^{\mu}$.\\

Now let's take a look at the parameter anisotropic factor $\Delta=p_t-p_r$, which measures the anisotropy where $p_t\neq p_r$. We have $T=\rho-p_r-2p_t$, which displays the trace of the anisotropic energy-momentum tensor. We can determine the geometry of wormholes based on the anisotropic factor. If $\Delta$ is negative ($\Delta<0$), the wormhole geometry is attractive, and the anisotropic force is directed inward. If $\Delta$ is positive ($\Delta>0$), the wormhole geometry is repulsive, and the anisotropic force is directed outward. The matter distribution of the wormholes exhibits isotropic pressure if $\Delta = 0$.\\

{In the context of Finsler gravity, the total action takes a similar form to that in Riemannian gravity by taking $R=Ric$. However, there are some key differences due to the nature of Finsler geometry.
	In the case of Finsler gravity, the total action can be expressed as follows \cite{Harko}
	\begin{equation}\label{eq.86}
		\mathcal{S}=\frac{1}{16\pi}\int d^x\sqrt{-g}f(Ric, T)+\int d^4x \sqrt{-g}\mathcal{L}_m
	\end{equation}
	where $\mathcal{L}_m$ is the matter Lagrangian density $g$ represents the determinant of the Finsler metric and $f(Ric, T)$ is the arbitrary function of $Ric$ and $T$
	Ricci scalar and energy-momentum tensor, respectively. In this section, we'll look at the functional form $f(Ric,T)= Ric+f(T)$, where $f(T)=2\lambda T$ is a function of the trace of the stress-energy tensor of matter. Our approach in deriving the modified gravitational field equation closely follows the methodology detailed in reference P. H. R. S. Moraes et al. \cite{Sahoo,Godani}. We obtained the Finslerian modified gravitational field equation as follows,
	\begin{equation}\label{eq.48}
		G^\mu_\nu=(8\pi_F+2\lambda)T^\mu_\nu+\lambda(2\rho+T)g^\mu_\nu,
	\end{equation}
	where $4\pi_F$ stands for the volume of the Finsler structure $\bar{F}$ in two dimensions. If we take $\bar{F}$ as the Finslerian sphere \cite{Li1}, then $\pi_F$ becomes equal to $\pi$. By combining components of the Finslerian modified Einstein tensor equation (\ref{eq.40}-\ref{eq.42}) and energy-momentum tensor equation (\ref{eq.47}), by taking  $(G=c=1)$ as relativistic units we can arrive at the gravitational field equations from equation (\ref{eq.48}) for Finslerian wormhole space-time equation (\ref{eq.27}) as follows \cite{Godani,Singh}},
\begin{eqnarray}
	G^t_t&=&(8\pi_F+2\lambda)T^t_t+\lambda(2\rho+T)g^t_t,  \nonumber \\
	\frac{1}{r^2}(b^{\prime}+\eta-1)&=&(8\pi_F+\lambda)\rho-\lambda(p_r+2p_t).\label{eq.49}\\
	G^r_r&=&(8\pi_F+2\lambda)T^r_r+\lambda(2\rho+T)g^r_r, ~~~\nonumber \\
	-\frac{b}{r^3}-\frac{1}{r^2}(\eta-1)&=&(8\pi_F+3\lambda)p_r+\lambda(\rho+2p_t).\label{eq.50}\\
	G^\mu_\mu &=&(8\pi_F+2\lambda)T^\mu_\mu+\lambda(2\rho+T)g^\mu_\mu ~~ (\mu=\theta,\phi), ~~\nonumber \\
	\frac{b-b^{\prime}r}{2r^3}&=&(8\pi_F+4\lambda)p_t+\lambda(\rho+p_r). \label{eq.51}
\end{eqnarray}
It is crucial to research how $\lambda$ affects wormhole solutions with shape functions in the presence of matter. If we know the distribution of the matter, gravitational field equations enable us to identify the space-time geometry and vice-versa. Additionally, it has the interesting property that, using the field equations, one can analyze both the distribution of matter and the whole structure of space-time if they are aware of some of the energy stress tensor's components and the space-time geometry.\\

\begin{figure}[hptb]
	\begin{center}
		\mbox{{\includegraphics[scale=0.4]{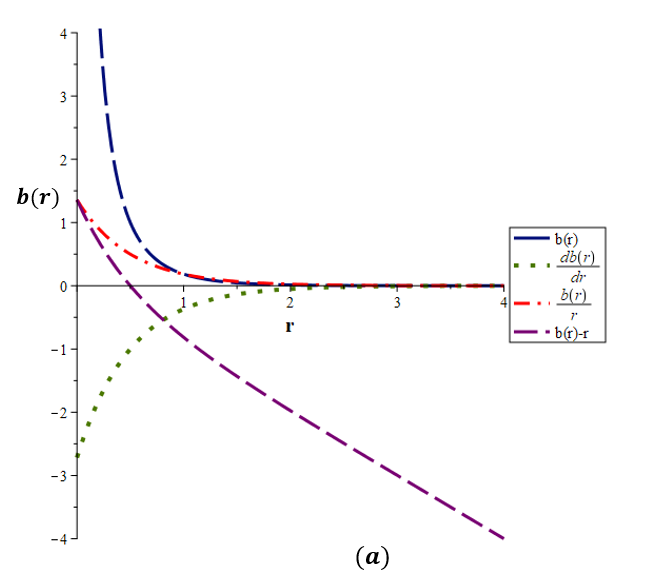}}
			{\includegraphics[scale=0.4]{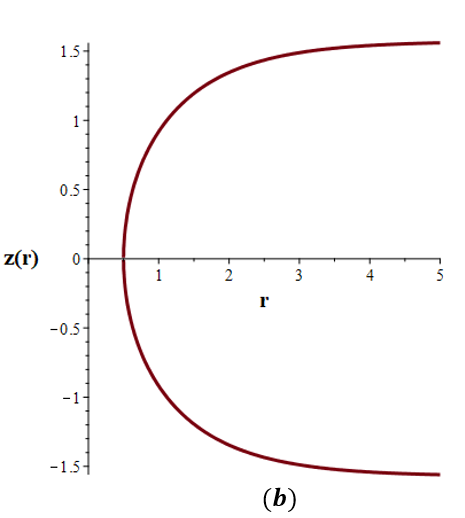}}}
		\caption{\label{fig1}  (a) Behavior of the exponential shape function equation (\ref{eq.52}) at $r_0=0.5$. (b) Embedded 2-D plot of the wormhole defined by equation (\ref{eq.52}).}
	\end{center}
\end{figure}

In the context of $f(R,T)=R +2\lambda T$ gravity, an exponential shape function~\cite{Moraes1}
\begin{equation}\label{eq.52}
	b(r)=r_0 e^{1-\frac{r}{r_0}},
\end{equation}
is taken into consideration in the Finslerian wormhole model to investigate the energy conditions. The shape function equation (\ref{eq.52}) meets every requirement for supporting wormholes mentioned above (Sec. \ref{sec2}), i.e., fulfills the requirements $r > r_0$, $b(r_0 )= r_0, 1-\frac{b(r)}{r} > 0$ where $r_0$  is the wormhole's throat radius, the flaring-out condition, and the asymptotic flatness condition. Fig. \ref{fig1}(a) depicts the properties of the exponential shape function equation (\ref{eq.52}) with $r_0= 0.5$ and demonstrates that the shape function meets all important requirements. We constructed an embedded 2-D diagram in fig. \ref{fig1}(b) and a 3-D diagram in fig. \ref{fig1a} for the shape function equation (\ref{eq.52}) for better visualization of the wormhole see Appendix.\\
\begin{figure}[hptb]
	\begin{center}
		
		{\includegraphics[scale=0.4]{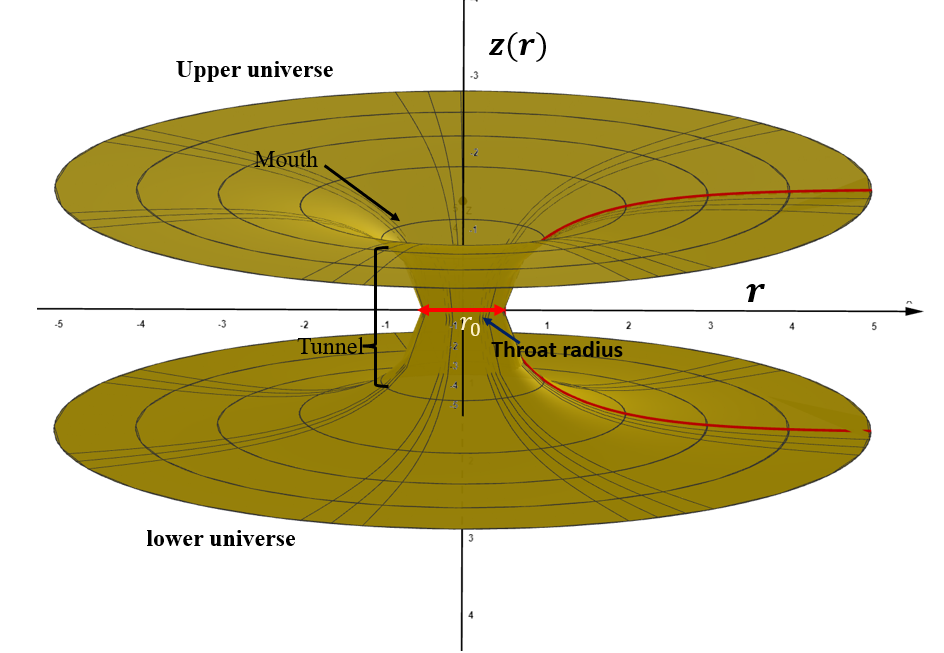}}
		
	\end{center}
	\caption{\label{fig1a}  Embedded 3-D plot of the wormhole defined by equation (\ref{eq.52}).}
\end{figure}
By inserting equation (\ref{eq.52}) into equations (\ref{eq.49}-\ref{eq.51}), energy density $\rho$, radial pressure $p_r$, and transverse pressure $p_t$ are obtained respectively as follows,
\begin{equation}\label{eq.53}
	\rho=\frac{-e^{1-\frac{r}{r_0}}+\eta-1}{r^2(8\pi_F+2\lambda)},
\end{equation}
\begin{equation}\label{eq.54}
	p_r=-\frac{r_0 e^{1-\frac{r}{r_0}}+r(\eta-1)}{r^3(8\pi_F+2\lambda)},
\end{equation}
\begin{equation}\label{eq.55}
	p_t=\frac{ e^{1-\frac{r}{r_0}}(r+r_0)}{2r^3(8\pi_F+2\lambda)}. \hspace{0.2cm}
\end{equation}
{fig. \ref{fig2}, fig. \ref{fig3}, and fig. \ref{fig4} depict the plot of $p_r$, $p_t$, $\rho$ respectively  as function of $r$ and $\eta$ for the Finslerian wormhole solution (\ref{eq.27}) at different value of $\lambda$, i.e., $8\pi_F\neq-2\lambda$ $(\lambda\neq-12.568)$  so we are taking $\lambda\geq-12.5$ (considering $\lambda=-12.5$) and  $\lambda \leq-12.6$ (considering $\lambda=-12.6$) with wormhole throat placed at $r_0=0.5$.}\\

\begin{figure}[hptb]
	\begin{center}
		\mbox{{\includegraphics[scale=0.32]{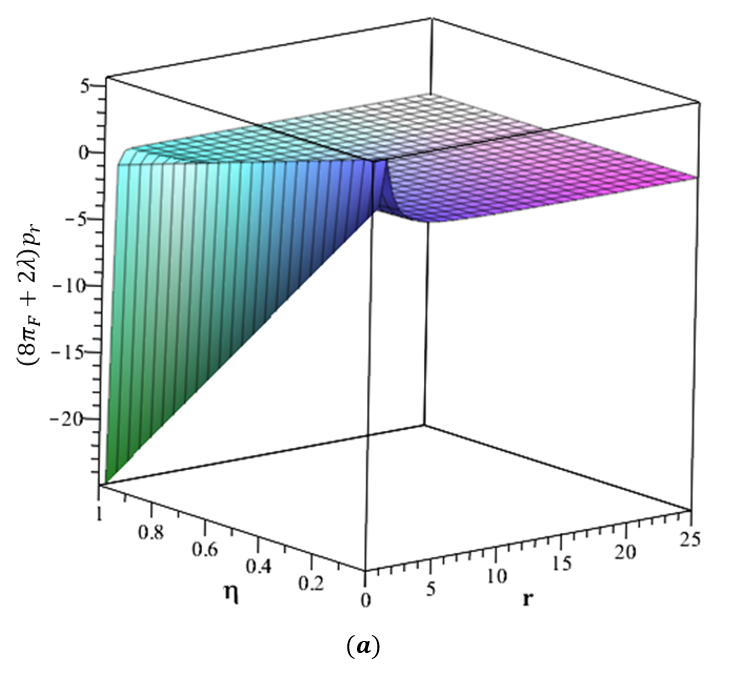}}
			{\includegraphics[scale=0.32]{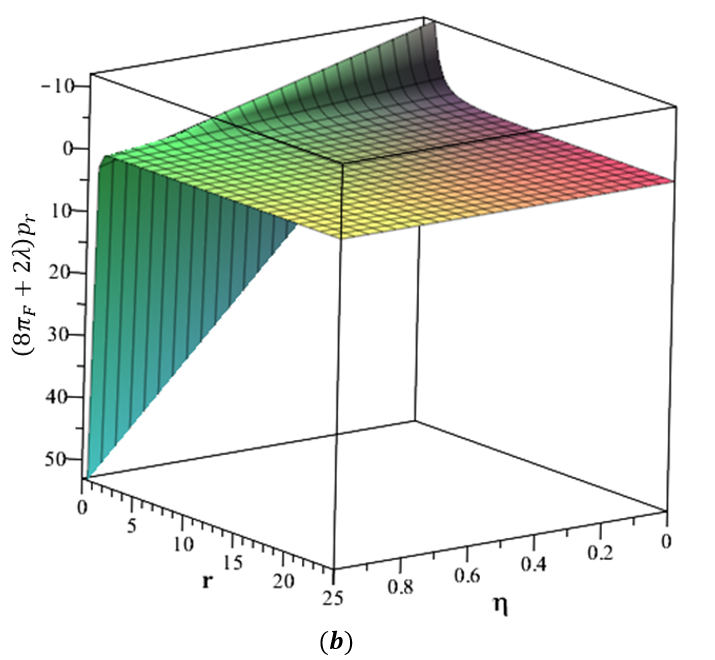}}}
		\caption{\label{fig2}  (a) Radial pressure $p_r=p_r(r, \eta)$ at $r_0=0.5$ when $\lambda\geq-12.5$. (b) Radial pressure $p_r=p_r(r, \eta)$ at $r_0=0.5$ when $\lambda\leq-12.6$.}
	\end{center}
\end{figure}
\begin{figure}[hptb]
	\begin{center}
		\mbox{{\includegraphics[scale=0.3]{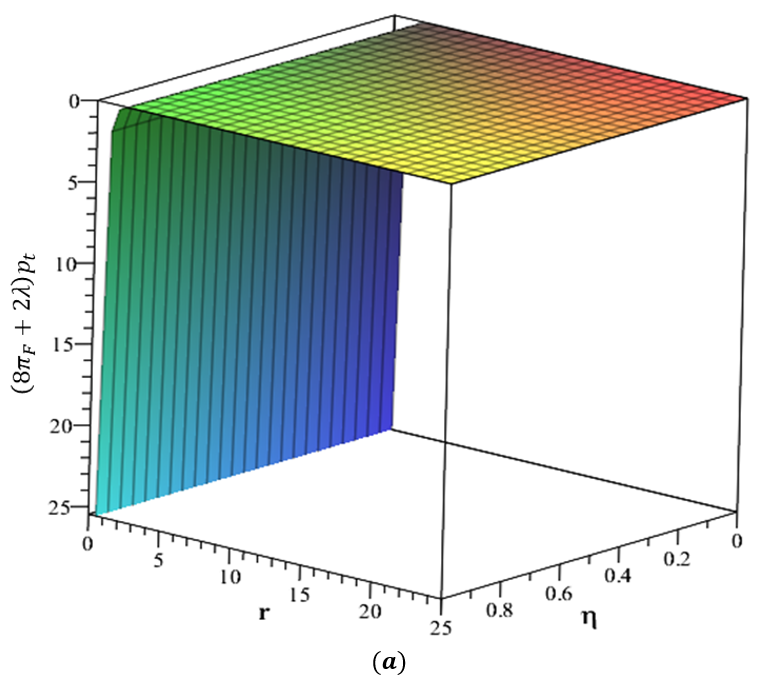}}
			{\includegraphics[scale=0.3]{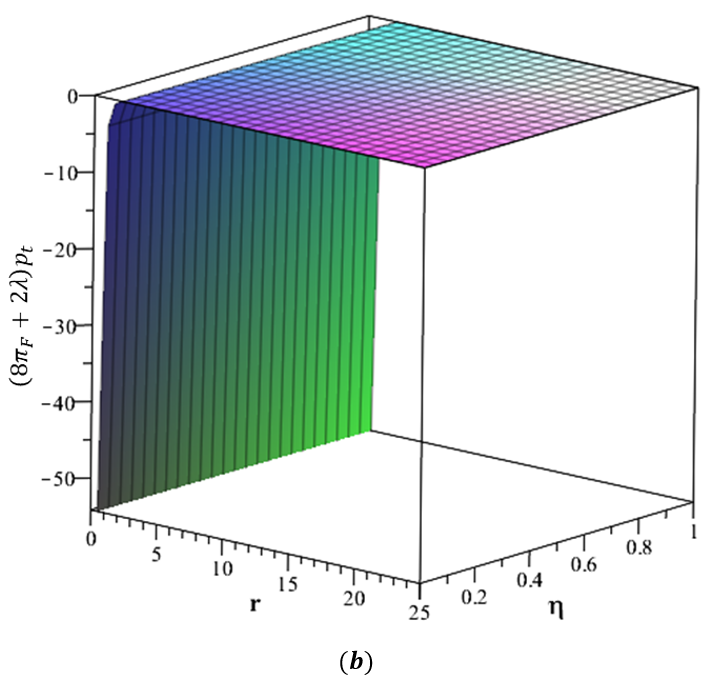}}}
		\caption{\label{fig3}  (a) Transverse pressure $p_t=p_t(r, \eta)$ at $r_0=0.5$ when $\lambda\geq-12.5$. (b) Transverse pressure $p_t=p_t(r, \eta)$ at $r_0=0.5$ when $\lambda\leq-12.6$.}
	\end{center}
\end{figure}
\begin{figure}[hptb]
	\begin{center}
		\mbox{{\includegraphics[scale=0.36]{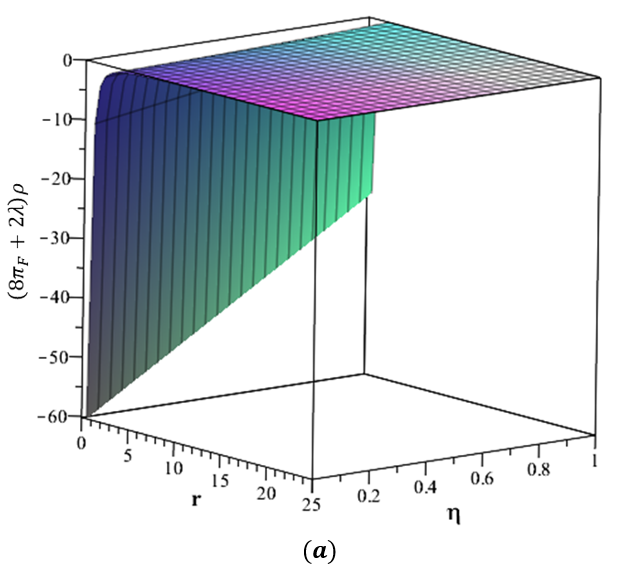}}
			{\includegraphics[scale=0.36]{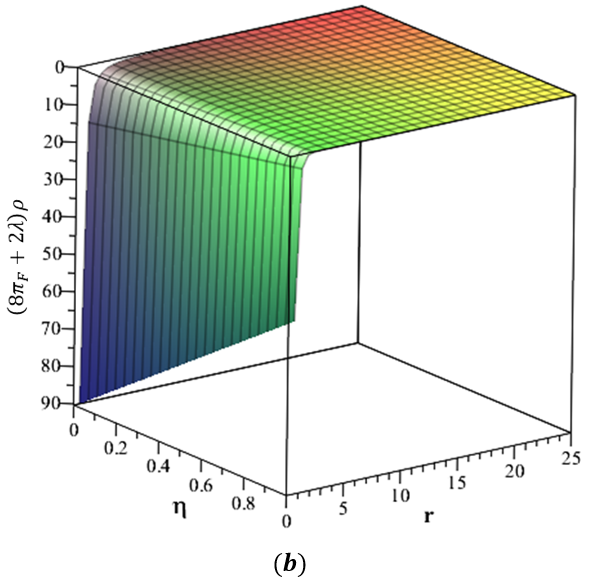}}}
		\caption{\label{fig4}  (a) Energy density $\rho=\rho(r, \eta)$ at $r_0=0.5$ when $\lambda\geq-12.5$ (WEC term). (b) Energy density $\rho=\rho(r, \eta)$ at $r_0=0.5$ when $\lambda\leq-12.6$ (WEC term).}
	\end{center}
\end{figure}
In addition, by combining the eqs. (\ref{eq.53}-\ref{eq.55}), we obtain the following stress-energy components,
\begin{equation}\label{eq.56}
	\rho+p_r=-\frac{e^{1-\frac{r}{r_0}}(r+r_0)}{r^3(8\pi_F+2\lambda)},
\end{equation}
\begin{equation}\label{eq.57}
	\rho+p_t=\frac{e^{1-\frac{r}{r_0}}(r_0-r)+2r(\eta-1)}{2r^3(8\pi_F+2\lambda)},
\end{equation}
\begin{equation}\label{eq.58}
	\rho+p_r+2p_t=0,
\end{equation}
\begin{equation}\label{eq.59}
	\rho-|p_r|=\frac{-e^{1-\frac{r}{r_0}}+\eta-1-r^2|\frac{r_0e^{1-\frac{r}{r_0}}+(\eta-1)r}{r^3}|}{r^2(8\pi_F+2\lambda)},
\end{equation}
\begin{equation}\label{eq.60}
	\rho-| p_t|=\frac{-2e^{1-\frac{r}{r_0}}+2\eta-2-|\frac{r_0+r}{r^3}| r^2e^{1-\frac{r}{r_0}}}{2r^3(8\pi_F+2\lambda)},
\end{equation}
\begin{equation}\label{eq.61}
	p_t-p_r=\frac{e^{1-\frac{r}{r_0}}(3r_0+r)+2r(\eta-1)}{2r^3(8\pi_F+2\lambda)},
\end{equation}
and
\begin{equation}\label{eq.62}
	\rho-p_r-2p_t=\frac{-e^{1-\frac{r}{r_0}}+\eta-1}{r^2(4\pi_F+\lambda)}.
\end{equation}
{The energy conditions, derived from the Raychaudhuri equations \cite{Raychaudhuri}, are crucial tools that involve inequalities among thermodynamic parameters like energy density ($\rho$), radial ($p_r$) and tangential pressures ($p_t$).
	These conditions, regardless of the theory of gravity, provide insights into the gravitational field's strength and play a key role in predicting the nature of space-time's geodesic structure. Particularly relevant for assessing the realism of matter distribution and potential wormhole existence, these conditions are purely geometrical concepts with broad applicability across gravitational theories}. The energy conditions are particularly effective tools for predicting the behavior of strong gravitational fields and the wormhole's geometry.
In GR, there are seven distinct kinds of energy conditions \cite{Visser} but we are just concentrating on four of them, such as Null energy condition (NEC), weak energy condition (WEC), strong energy condition (SEC), and dominant energy condition (DEC).
These energy conditions are characterized in terms of $\rho$,  $p_r$, and  $p_t$.
\begin{itemize}
	\item According to the WEC, any time-like observer's measurement of energy density must be always positive, i.e., $\rho \geq 0$, and for all $l$,   $\rho + p_l\geq 0$.
	\item According to the SEC, gravity should always be attractive. Regarding the components of the energy-momentum tensor, i.e., for all $l$, $\rho + p_l\geq 0$ and $\rho+ \sum_l p_l\geq 0$.
	\item The DEC states that any observer's measurement of the energy flux is null or time-like and should be non-negative, i.e., $\rho\geq 0$, which results in $\rho\geq | p_l|$.
	\item For SEC and WEC basic requirement is NEC, i.e., for all $l$,  $\rho + p_l\geq 0$.\\ All of the above-mentioned energy conditions are implied to be invalid by the NEC violation.
\end{itemize}
The NEC follows from the SEC, but the reverse need not be true. Moreover, the SEC need not illustrate the WEC. The DEC implies both the NEC and the WEC, while the reverse doesn't need to be true in each case. Furthermore, the DEC need not imply the SEC. The SEC \cite{Mishra} is a result of gravity's attractive characteristics, and its shape originates directly from an analysis of a spherically symmetrical metric in the GR system.
To describe the geometry of wormholes, the SEC violation is essential. Although the violation of energy requirements is quite acceptable in some quantum fields, the violation of NEC (commonly referred to as exotic matter) is a basic aspect of static traversable wormholes in the context of GR. Due to this NEC violation, all of the standard energy conditions were also violated.\\

\begin{figure}[hptb]
	\begin{center}
		\mbox{{\includegraphics[scale=0.35]{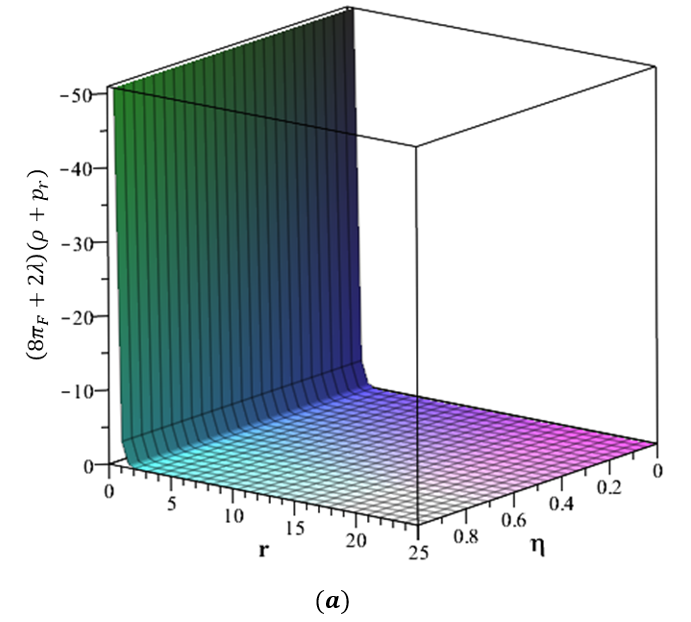}}
			{\includegraphics[scale=0.38]{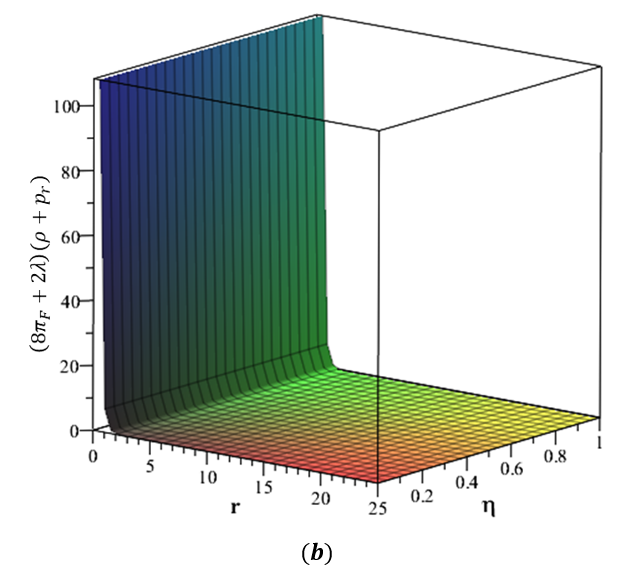}}}
		\caption{\label{fig5}  (a) $\rho+p_r=(\rho+p_r)(r, \eta)$ at $r_0=0.5$ when $\lambda\geq-12.5$ ( NEC term). (b) $\rho+p_r=(\rho+p_r)(r, \eta)$ at $r_0=0.5$ when $\lambda\leq-12.6$ (NEC term).}
	\end{center}
\end{figure}

\begin{figure}[hptb]
	\begin{center}
		\mbox{{\includegraphics[scale=0.36]{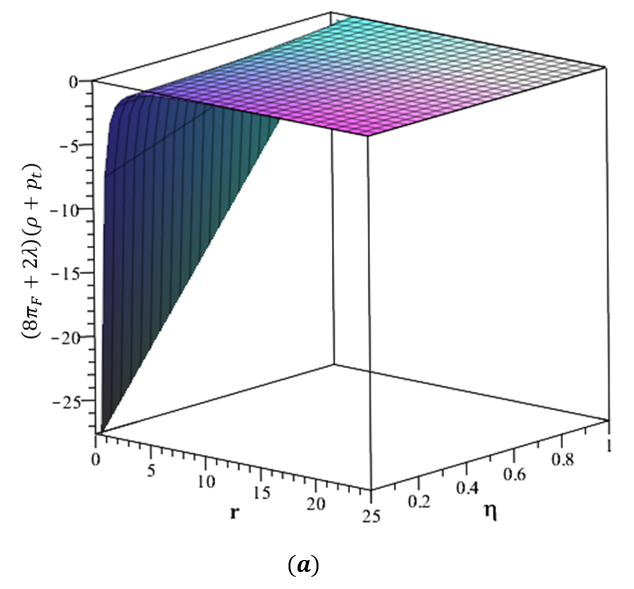}}
			{\includegraphics[scale=0.36]{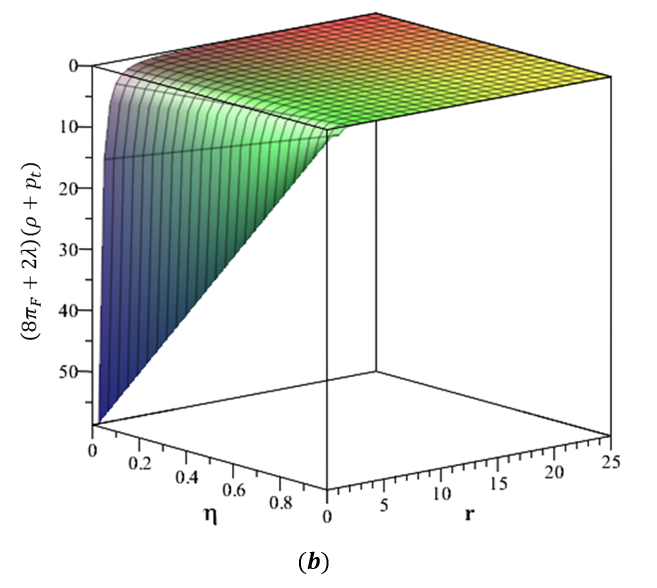}}}
		\caption{\label{fig6}  (a) $\rho+p_t=(\rho+p_t)(r, \eta)$ at $r_0=0.5$ when $\lambda\geq-12.5$ (NEC term). (b) $\rho+p_t=(\rho+p_t)(r, \eta)$ at $r_0=0.5$ when $\lambda\leq-12.6$ (NEC term).}
	\end{center}
\end{figure}

\begin{figure}[hptb]
	\begin{center}
		\mbox{{\includegraphics[scale=0.35]{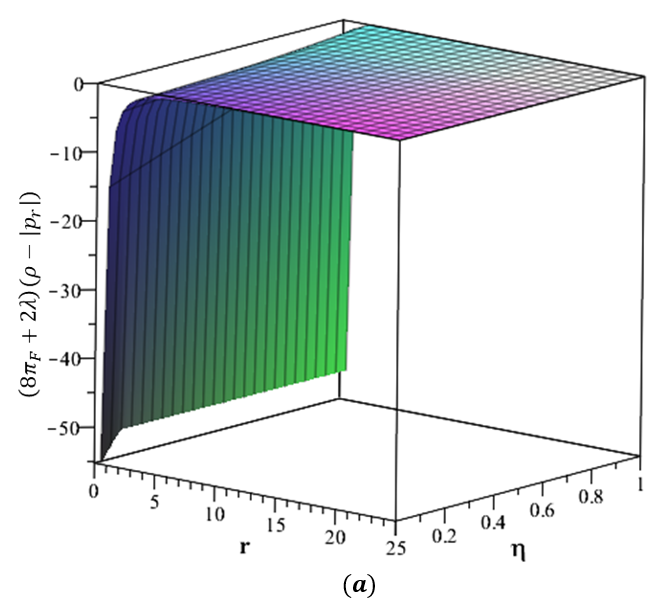}}
			{\includegraphics[scale=0.37]{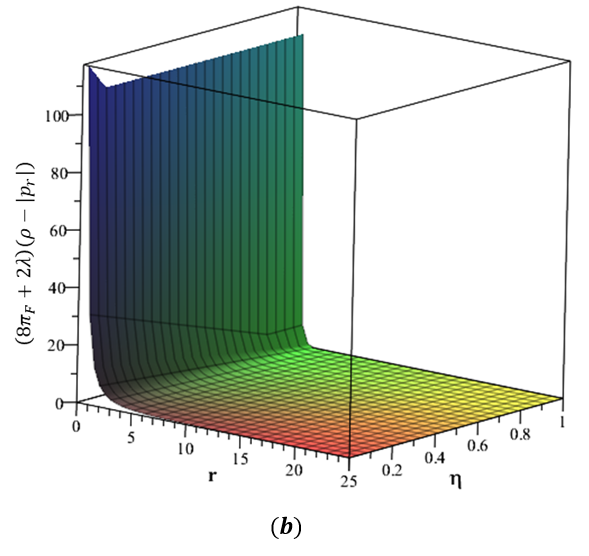}}}
		\caption{\label{fig7}  (a) $\rho-| p_r|=(\rho-| p_r|)(r, \eta)$ at $r_0=0.5$ when $\lambda\geq-12.5$ (DEC term). (b) $\rho-| p_r|=(\rho-| p_r|)(r, \eta)$ at $r_0=0.5$ when $\lambda\leq-12.6$ (DEC term).}
	\end{center}
\end{figure}

\begin{figure}[hptb]
	\begin{center}
		\mbox{{\includegraphics[scale=0.35]{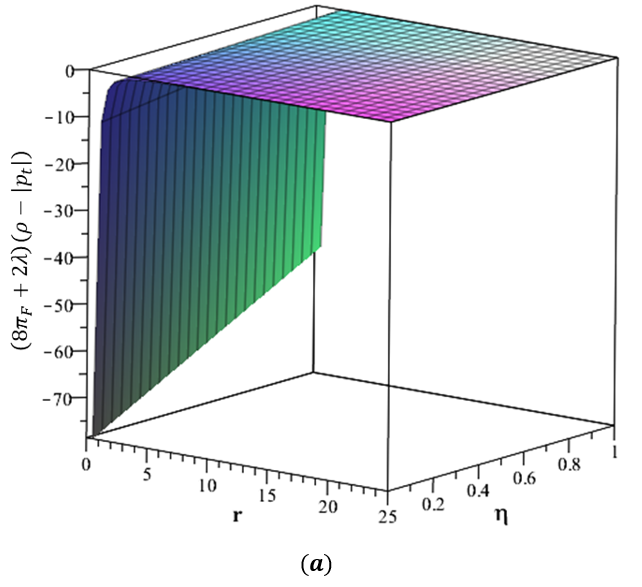}}
			{\includegraphics[scale=0.35]{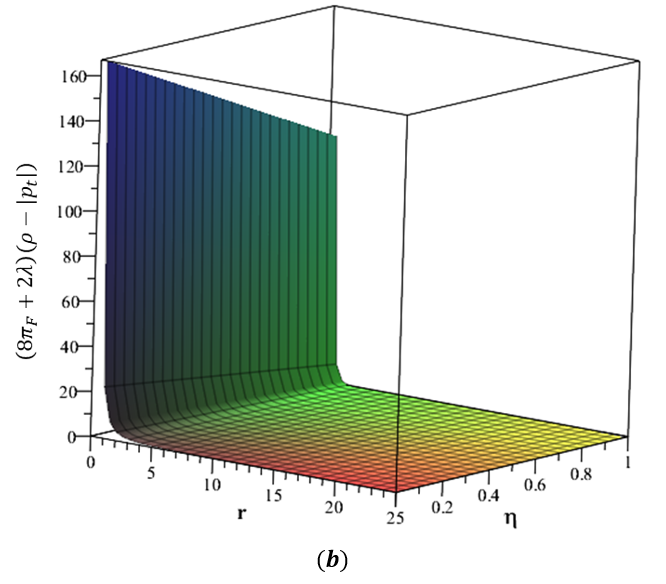}}}
		\caption{\label{fig8}  (a) $\rho-| p_t|=(\rho-| p_t|)(r, \eta)$ at $r_0=0.5$ when $\lambda\geq-12.5$ (DEC term). (b)  $\rho-| p_t|=(\rho-| p_t|)(r, \eta)$ at $r_0=0.5$ when $\lambda\leq-12.6$ (DEC term).}
	\end{center}
\end{figure}

\begin{figure}[hpbt]
	\begin{center}
		\includegraphics[scale=0.43]{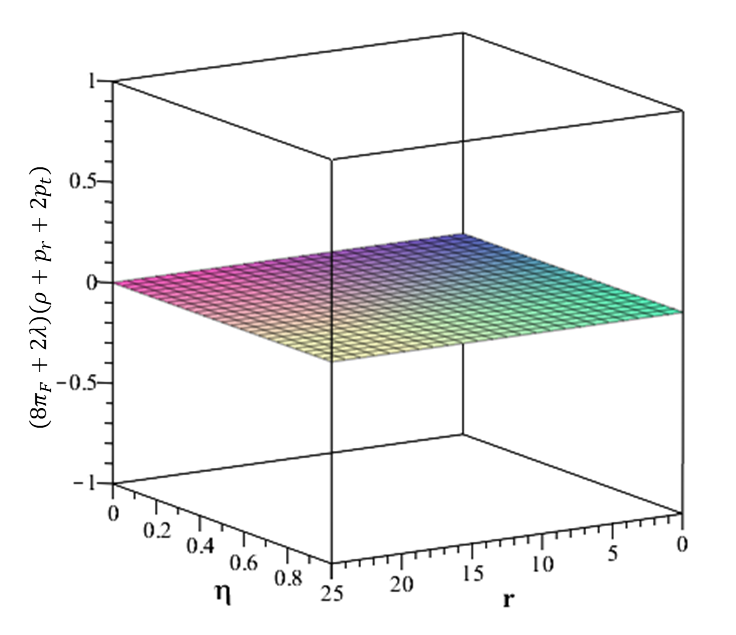}
		\caption{\label{fig9}$\rho+p_r+2p_t=(\rho+p_r+2p_t)(r, \eta)$ at $r_0=0.5$ (SEC term).}
	\end{center}
\end{figure}

\begin{figure}[hptb]
	\begin{center}
		\mbox{{\includegraphics[scale=0.35]{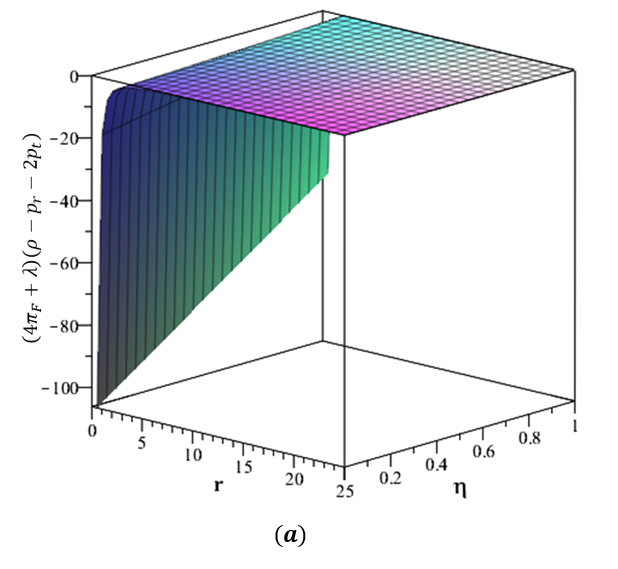}}
			{\includegraphics[scale=0.34]{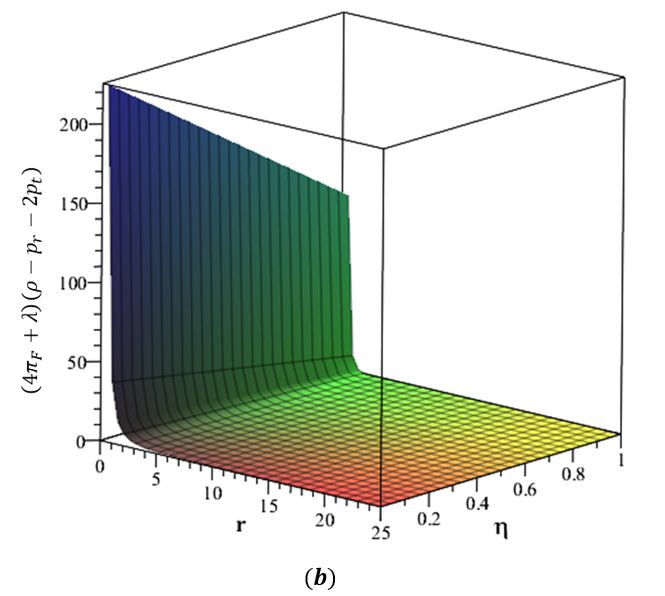}}}
		\caption{\label{fig10}  (a) $T=\rho-p_r-2p_t=(\rho-p_r-2p_t)(r, \eta)$ at $r_0=0.5$ when $\lambda\geq-12.5$ (SEC term). (b) $T=\rho-p_r-2p_t=(\rho-p_r-2p_t)(r, \eta)$ at $r_0=0.5$ when $\lambda\leq-12.6$ (SEC term).}
	\end{center}
\end{figure}

\begin{figure}[hptb]
	\begin{center}
		\mbox{{\includegraphics[scale=0.34]{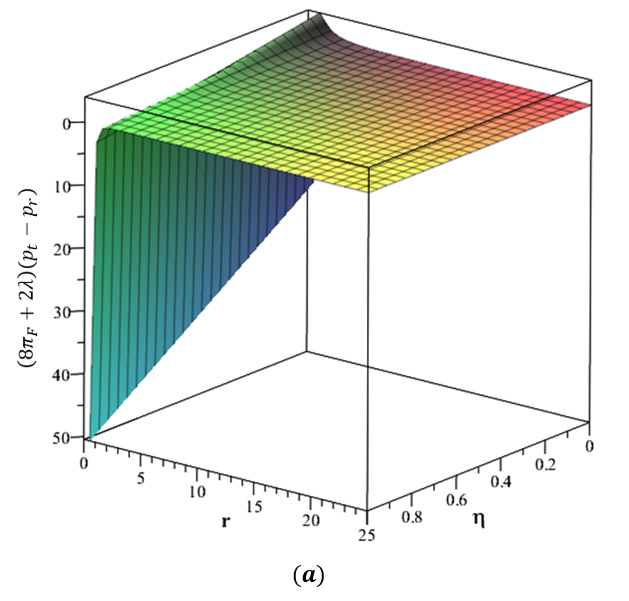}}
			{\includegraphics[scale=0.34]{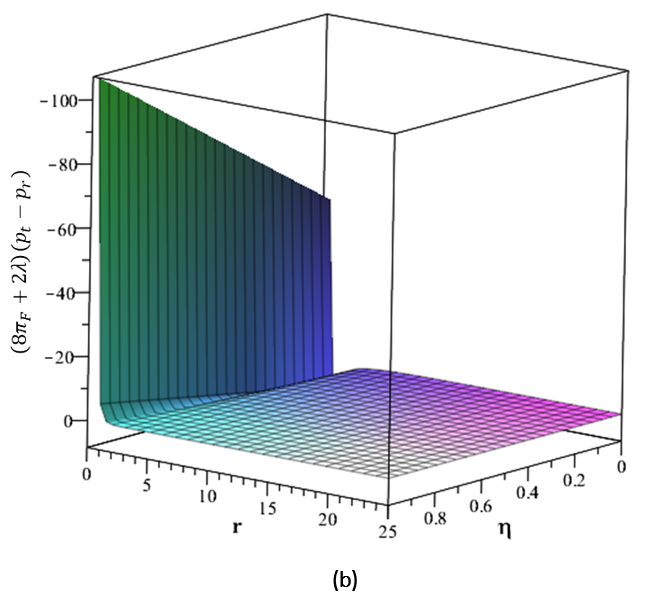}}}
		\caption{\label{fig11}  (a) $\Delta=\Delta(r, \eta)$ the anisotropic factor at $r_0=0.5$ when $\lambda\geq-12.5$. (b) $\Delta=\Delta(r, \eta)$ the anisotropic factor at $r_0=0.5$ when $\lambda\leq-12.6$.}
	\end{center}
\end{figure}

From figs. (\ref{fig4}-\ref{fig11}), we plotted the four energy conditions and anisotropic factor for different values of $\lambda$ as a function of $r$  and $\eta$ with wormhole throat at $r_0=0.5$ for the current Finslerian wormhole model in $f(Ric,T)$ MGT.\\

\begin{figure}[hptb]
	\begin{center}
		\mbox{{\includegraphics[scale=0.43]{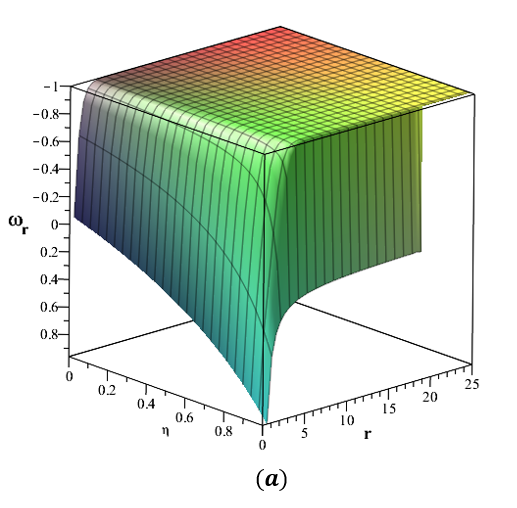}}
			{\includegraphics[scale=0.43]{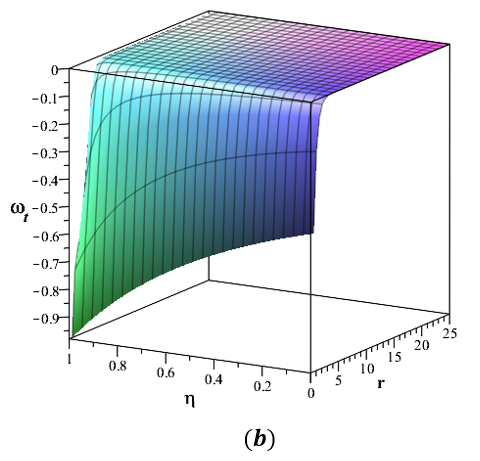}}}
		\caption{\label{fig12}  (a) $\omega_r=\omega_r(r, \eta)$ radial EoS parameter at $r_0=0.5$. (b) $\omega_t=\omega_t(r, \eta)$ transverse EoS parameter  at $r_0=0.5$.}
	\end{center}
\end{figure}

We will need barotropic EoS for the radial and transverse pressures. we express it by
\begin{equation}\label{eq.63}
	p_r=\omega_r\rho, \hspace{1cm} p_t=\omega_t\rho,
\end{equation}
where $\omega_t$ and $\omega_r$ stand for the transverse and radial EoS parameters, respectively. SEC is specified by equation (\ref{eq.58}), which states that the inhomogeneous and anisotropic component fulfills. In this specific instance, $-1\leq\omega_r\leq1$ followed by $-1\leq\omega_t\leq0$, along with $\omega_r-1\geq0$ and $\omega_r+1\leq0$, and we observed that $\omega_t\leq-1$ and $\omega_t\geq0$ respectively. From equation (\ref{eq.63}) $\omega_r$ and $\omega_t$ can be written by using eqs. (\ref{eq.53}-\ref{eq.55}) as follows,
\begin{equation}\label{eq.64}
	\omega_r=\frac{r_0 e^{1-\frac{r}{r_0}}+(\eta-1)r}{(e^{1-\frac{r}{r_0}}-\eta+1)r},
\end{equation}
\begin{equation}\label{eq.65}
	\omega_t=\frac{ e^{1-\frac{r}{r_0}}(r_0+r)}{(-e^{1-\frac{r}{r_0}}+\eta-1)2r}.
\end{equation}
The following expression is satisfied by the barotropic EoS parameters $\omega_r$ and $\omega_t$
\begin{equation}\label{eq.66}
	\omega_r+2\omega_t+1=0.
\end{equation}
With reference to the radial EoS parameter $\omega_r$, the transverse pressure $p_t$ can be expressed as follows,
\begin{equation}\label{eq.67}
	p_t=-\frac{1}{2}(1+\omega_r)\rho.
\end{equation}
It is evident that $1 +\omega_r\geq 0$ from \ref{fig12}(a). As a result, equation (\ref{eq.67}) suggests that $p_t\geq0$. Accordingly, fig. \ref{fig3} makes this point clear.
In accordance with the indicated range of radial coordinate $r$ and the  $\eta$,  with the wormhole throat at $r_0=0.5$, the radial EoS parameter $\omega_r$ and the transverse EoS parameter $\omega_t$ are shown in fig. \ref{fig12}.

Now we are considering $F_a$ stands for the anisotropic wormhole force, $F_h$ for the hydrostatic force, and $F_g$ for the gravity force. The formulas for $F_a, F_h$ and $F_g$ are provided by
\begin{equation}\label{eq.68}
	F_a=\frac{2(p_t-p_r)}{r},
\end{equation}
\begin{equation}\label{eq.69}
	F_h=-p'_r, \hspace{1cm}
\end{equation}
\begin{equation}\label{eq.70}
	F_g=-a'(r)(\rho+p_r).\hspace{-0.6cm}
\end{equation}
The following is the conservation equation
\begin{equation}\label{eq.71}
	F_a+F_h+F_g=0.
\end{equation}
The equations for $F_a, F_h$, and $F_g$ for the Finslerian wormhole space-time (\ref{eq.27}) are obtained from eqs. (\ref{eq.53}-\ref{eq.55}), and are used in eqs. (\ref{eq.68}-\ref{eq.70}).
\begin{equation}\label{eq.72}
	F_a=\frac{e^{1-\frac{r}{r_0}}(3r_0+r)+(\eta-1)2r}{r^4(8\pi_F+2\lambda)},
\end{equation}
\begin{equation}\label{eq.73}
	F_h=-\frac{e^{1-\frac{r}{r_0}}(3r_0+r)+(\eta-1)2r}{r^4(8\pi_F+2\lambda)},
\end{equation}
\begin{equation}\label{eq.74}
	F_g=a'(r)\left(\frac{e^{1-\frac{r}{r_0}}(r+r_0)}{r^3(8\pi_F+2\lambda)}\right)=0.\hspace{.5cm}
\end{equation}
Since we are studying tideless wormholes, we taken $a(r)=$ constant implies $a^{\prime}(r)=0$. For this condition, the force due to the contribution of gravity equation (\ref{eq.74}) becomes zero in our study model. The tidal gravitational forces that travelers experience must be minimal for the wormhole to be traversable. So the condition $a^{\prime}(r)=0$ supports the features of a traversable wormhole. Because of this conservation equation equation (\ref{eq.71}), becomes
\begin{equation}\label{eq.75}
	F_a=-F_h.
\end{equation}
We deduce that the Finslerian wormhole solution is in equilibrium because our observation in equation (\ref{eq.75}) shows that the opposing effects of hydrostatic and anisotropic forces cancel each other.

\begin{figure}[hptb]
	\begin{center}
		\mbox{{\includegraphics[scale=0.34]{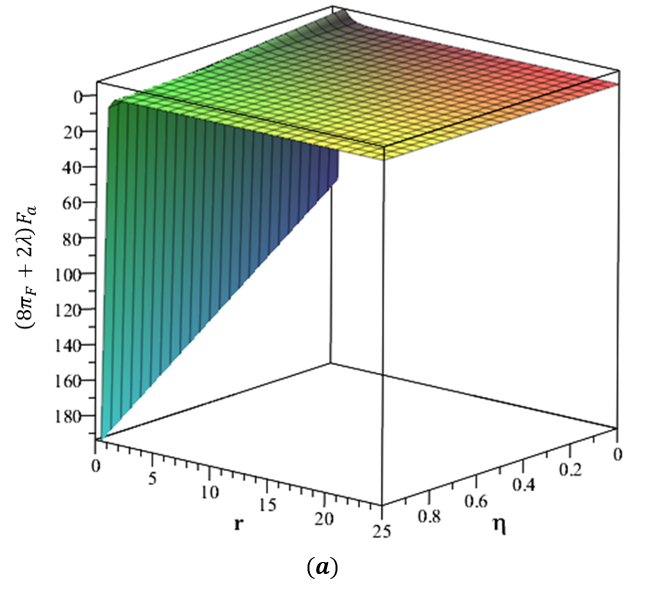}}
			{\includegraphics[scale=0.34]{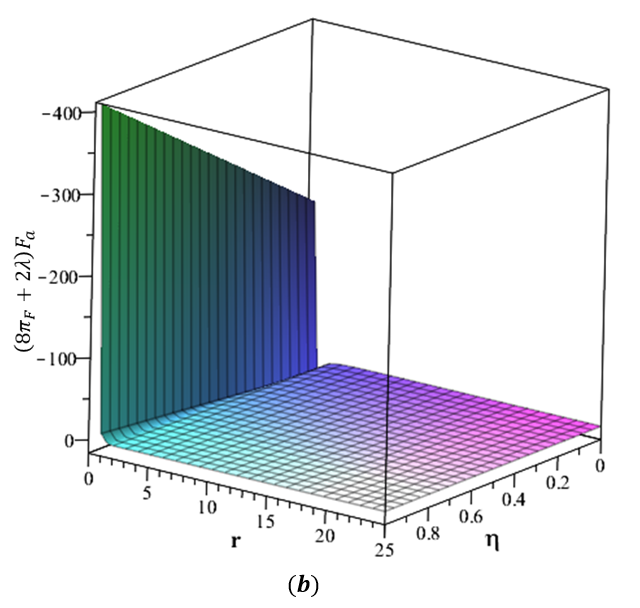}}}
		\caption{\label{fig13}  (a) $F_a=F_a(r, \eta)$ variation of anisotropic force at $r_0=0.5$ when $\lambda\geq-12.5$. (b) $F_a=F_a(r, \eta)$ variation of anisotropic force at $r_0=0.5$ when $\lambda\leq-12.6$.}
	\end{center}
\end{figure}

\begin{figure}[hptb]
	\begin{center}
		\mbox{{\includegraphics[scale=0.47]{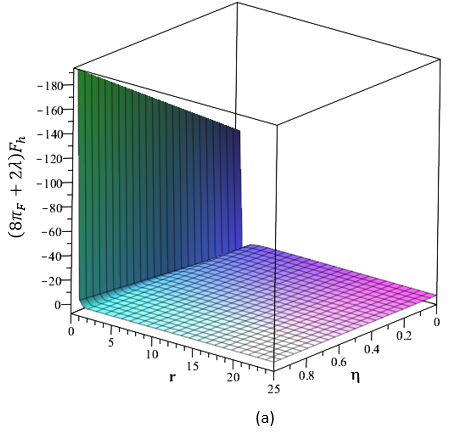}}
			{\includegraphics[scale=0.47]{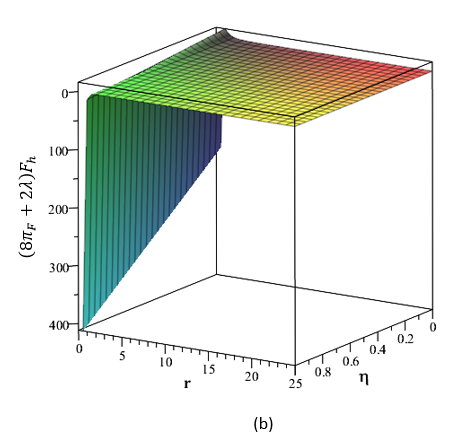}}}
		\caption{\label{fig14}  (a) $F_h=F_h(r, \eta)$ variation of hydrostatic force at $r_0=0.5$ when $\lambda\geq-12.5$. (b) $F_h=F_h(r, \eta)$ variation of hydrostatic force at $r_0=0.5$ when $\lambda\leq-12.6$.}
	\end{center}
\end{figure}

In figs. \ref{fig13} and \ref{fig14}, the variation of the hydrostatic force $F_h$, and the anisotropic force $F_a$ against the radial coordinate $r$ with the change in the $\eta$ and at $r_0 = 0.5$ shown for distinct  values of $\lambda$.

\begin{figure}[hptb]
	\begin{center}
		\mbox{{\includegraphics[scale=0.33]{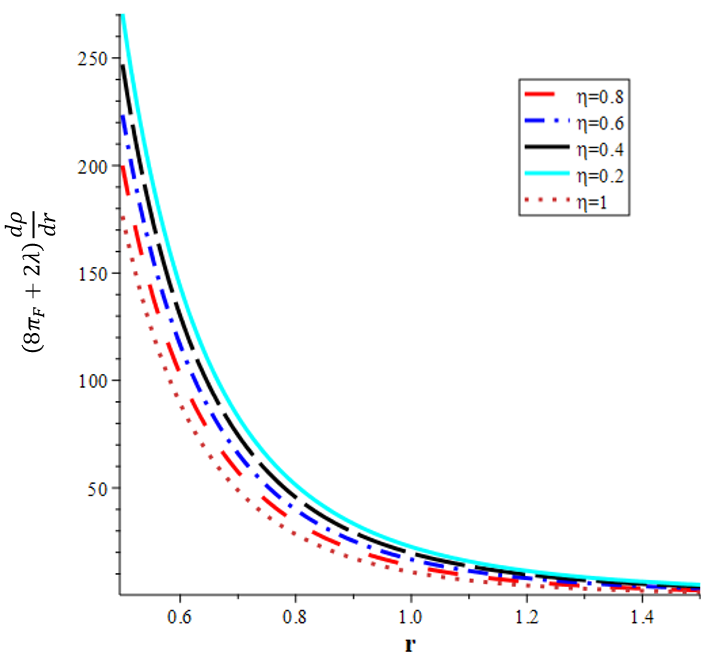}}
			{\includegraphics[scale=0.33]{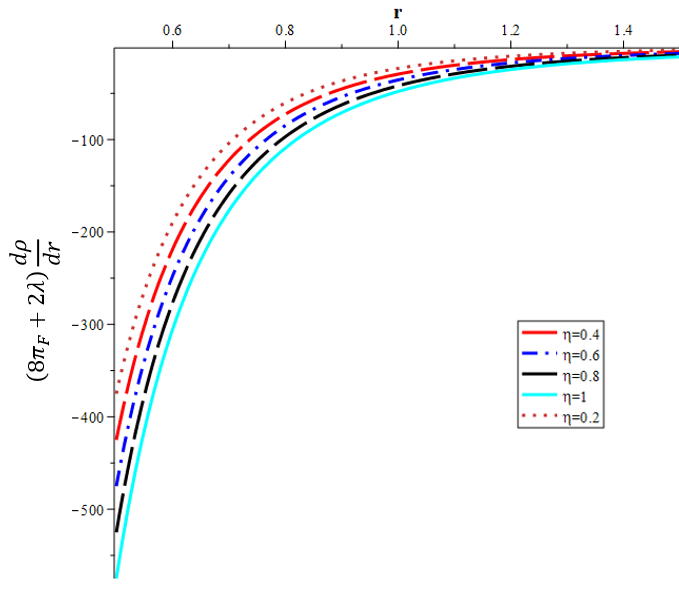}}}
		\caption{\label{fig15}  (a) Variation of $\frac{d\rho}{dr}=\frac{d\rho}{dr}(r, \eta)$ at  $r_0=0.5$ when $\lambda\geq-12.5$. (b) Variation of $\frac{d\rho}{dr}=\frac{d\rho}{dr}(r, \eta)$ at  $r_0=0.5$ when $\lambda\leq-12.6$.}
	\end{center}
\end{figure}

\begin{figure}[hptb]
	\begin{center}
		\mbox{{\includegraphics[scale=0.35]{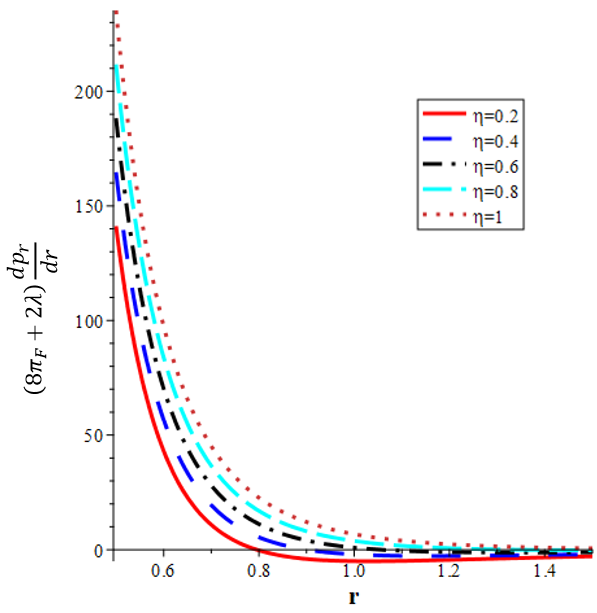}}
			{\includegraphics[scale=0.35]{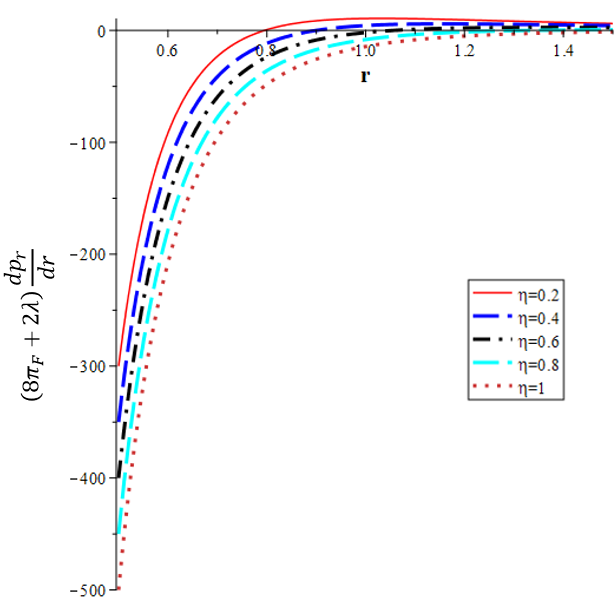}}}
		\caption{\label{fig16}  (a) Variation of $\frac{dp_r}{dr}=\frac{dp_r}{dr}(r, \eta)$ at  $r_0=0.5$ when $\lambda\geq-12.5$. (b) Variation of $\frac{dp_r}{dr}=\frac{dp_r}{dr}(r, \eta)$ at  $r_0=0.5$ when $\lambda\leq-12.6$.}
	\end{center}
\end{figure}

\begin{figure}[hpbt]
	\begin{center}
		\includegraphics[scale=0.5]{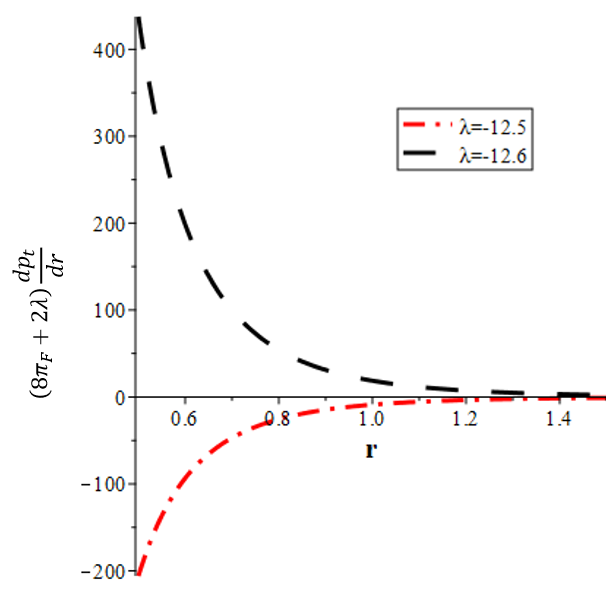}
		\caption{\label{fig17} Variation of $\frac{dp_t}{dr}=\frac{dp_t}{dr}(r)$ at  wormhole throat $r_0=0.5$.}
	\end{center}
\end{figure}

Considering the following formulas are used to express the energy density gradient $\frac{d\rho}{dr}$, radial pressure gradient $\frac{dp_r}{dr}$, and transverse pressure gradient $\frac{dp_t}{dr}$, respectively,
\begin{equation}\label{eq.76}
	\frac{d\rho}{dr}=\frac{e^{1-\frac{r}{r_0}}(r+2r_0)-(\eta-1)2r_0}{r_0 r^3(8\pi_F+2\lambda)},
\end{equation}
\begin{equation}\label{eq.77}
	\frac{dp_r}{dr}=\frac{e^{1-\frac{r}{r_0}}(r+3r_0)+(\eta-1)2r}{r^4(8\pi_F+2\lambda)},
\end{equation}
\begin{equation}\label{eq.78}
	\frac{dp_t}{dr}=-\frac{e^{1-\frac{r}{r_0}}(r^2+3rr_0+3r^2_0)}{2r_0r^4(8\pi_F+2\lambda)}.
\end{equation}
Using equation (\ref{eq.74}-\ref{eq.76}) we have following,
\begin{equation}\label{eq.79}
	\frac{dp_r}{d\rho}= \frac{r_0 e^{1-\frac{r}{r_0}}(r+3r_0)+(\eta-1)2rr_0}{re^{1-\frac{r}{r_0}}(r+2r_0)-(\eta-1)2rr_0},
\end{equation}
\begin{equation}\label{eq.80}
	\frac{dp_t}{d\rho}=-\frac{e^{1-\frac{r}{r_0}}(r^2+3rr_0+3r^2_0)}{2[re^{1-\frac{r}{r_0}}(r+2r_0)-(\eta-1)2rr_0]}.
\end{equation}
Showing the  gradient $\frac{d\rho}{dr}$, $\frac{dp_r}{dr}$, and $\frac{dp_t}{dr}$,  how they are varying in figs. (\ref{fig15}-\ref{fig17}), respectively, for different value of $\lambda$ that change with $r$ for given value of $\eta$ for the wormhole throat at $r_0= 0.5$.
\section{Results and  discussion}\label{sec4}
Modified theories have the flexibility to experiment with the effective stress-energy tensor to find an alternative approach for solving the exotic matter problem while still adhering to all energy conditions. $f(Ric,T)$ gravity is one of the modified theories of gravity that has received a lot of interest. With an exponential shape function that corresponds to the family $f(Ric, T)$ of extended theories of gravity, we have developed a Finslerian wormhole model in our present study. We focused our studies to the case $f(Ric,T)=Ric+f(T)$, where $f(T)=2\lambda T$ and $\lambda$ is constant parameter.  {The nature of matter composition in $f(Ric,T)$ gravity wormholes is strongly influenced by parameter choices, with the study by Chanda et al. \cite{Chanda} emphasizing the significant impact of the parameter $\lambda$.}  The equilibrium configurations of white dwarfs are investigated by  G. A. Carvalho et al. \cite{Carvalho} using an MGT that depends on the parameter $\lambda$. As a result, numerous physicists and geometrists have this generalized geometry as interesting. Consequently, the Finslerian space-time manifold is of significance to us when we study the wormhole configuration in $f(Ric,T)$ MGT. In general, the "exotic matter" violates the weak/null energy conditions and has the unique capability to keep the wormhole tunnel open. We can reveal that NEC is an artifact of the GR that supports a wormhole. A key requirement of static traversable wormholes, according to GR, is the violation of different energy conditions, often known as exotic matter.\\

In the current work, fig. \ref{fig1} shows that the exponential shape function equation (\ref{eq.52}) meets all essential geometric conditions, and as a result, we obtained the structure of the wormhole. We plotted WEC, NEC, DEC, and SEC energy conditions in figs. (\ref{fig4}-\ref{fig11}). We plotted WEC in fig. \ref{fig4} and observed that in fig. \ref{fig4}(a) energy density condition is negative and also tends to zero with the change of radial coordinate $r$ with  $\eta$ at $r_0=0.5$, for the value  $\lambda\geq-12.5$ and in fig. \ref{fig4}(b) energy density condition is positive and also tends to zero with the change of radial coordinate $r$ with  $\eta$ at $r_0=0.5$, for the value $\lambda\leq-12.6$. Therefore, we conclude that WEC is violated in the Finslerian wormhole $f(Ric,T)$ gravity model for the value $\lambda\geq-12.5$, within the specified range of $r$ and $\eta$.\\

In the figs. (\ref{fig5}-\ref{fig8}) we plotted the energy condition terms, using the specified range of $r$ and $\eta$ at $r_0=0.5$, and found $\rho+p_r$ in fig. \ref{fig5}, $\rho+p_t$ in fig. \ref{fig6}, $\rho-|p_r|$ in fig. \ref{fig7}, and $\rho-|p_t|$ in fig. \ref{fig8}, negative for the value of $\lambda\geq-12.5$ and  positive for the value of $\lambda\leq-12.6$. With equation (\ref{eq.56}), we determined that the $\eta$ is not a factor in the NEC term $\rho+p_r$, which is the function of $r$ only. But we have demonstrated the term $\rho+ p_r$ graphically in a 3-D plot in fig. \ref{fig5} and discovered that the result is negative for the value $\lambda\geq-12.5$ in fig. \ref{fig5}(a), and positive for the value $\lambda\geq-12.6$ in fig. \ref{fig5}(b). From fig. \ref{fig6}(a) we observed $\rho+p_t$ is negative for $\lambda\geq-12.5$, and positive for $\lambda\leq-12.6$. As a result, we conclude NEC is violated in Finslerian $f(Ric, T)$ gravity at $\lambda\geq-12.5$, hence Finslerian wormhole to be traversable. Also, we observed DEC in fig. \ref{fig8}(a) violets for the value $\lambda\geq-12.5$. From these findings, we draw the following conclusions in the framework of Finslerian $f(Ric,T)$ gravity the wormholes must be filled with exotic matter.\\

From equation (\ref{eq.58})  in fig. \ref{fig9} $\rho+p_r+p_t$ is equal to zero for the given range of $r$ and $\eta$, and we observed the violation of SEC. The $\eta$ determines whether the Finslerian wormhole is attractive or repulsive. According to fig. \ref{fig11}(a), the Finslerian wormhole is repulsive for $1 \geq\eta\geq0.8$, it is initially repulsive and then becomes attractive after crossing the certain value of $r$ for $0.2 \leq\eta< 0.8$ for $\lambda\geq-12.5$, and it is opposite when $\lambda\leq-12.6$ in fig. \ref{fig11}(b). We observed from fig. \ref{fig12}(a) that the radial EoS parameter $\omega_r$ takes the values between -1 and 1 for the specified value of $r$ and $\eta$. Further, if we replace  $\bar{F}$ with the Finslerian sphere, then $\omega_r$ is obtained to be non-negative with values less than 1. The transverse EoS parameter $\omega_t$ lies between the value of -1 and 0 in fig. \ref{fig12}(b)  because of equation (\ref{eq.58}) and $-1\leq\omega_r\leq 1$ \cite{Cataldo}. The wormhole matter distribution is affected by three forces because of anisotropic pressure. The Finslerian wormhole solution is in equilibrium as a result of the combined effect of the anisotropic force $F_a$ and hydrostatic force $F_h$ in fig. \ref{fig13}(a) and fig. \ref{fig14}(a) respectively, for $\lambda\leq-12.5$. In the beginning, the anisotropic force and the hydrostatic force have positive values for the range $1\geq\eta\geq 0.6$ with a variation of $r$ and at the value of $\lambda\geq-12.5$, and negative values for the value of $0.2\leq\eta< 0.6$ respectively. As a result, they are repulsive and attractive. Their nature changes as $r$ reach a certain value.\\

The energy density gradient $\frac{d\rho}{dr}$ is positive in fig. \ref{fig15}(a)
for $\lambda\geq-12.5$ and negative in fig. \ref{fig15}(b) for $\lambda\leq-12.6$, with change of $r$ for each value of the  $\eta$ at $r_0=0.5$. We can observe from fig. \ref{fig16}(a) that the radial pressure gradient $\frac{dp_r}{dr}$ is found to be dependent on the value of $\eta$. $\frac{dp_r}{dr}$ is positive For $\eta= 0.6, 0.8$ and 1, for distinct values of $r$. But $\frac{dp_r}{dr}$ is positive for
$\eta=0.2$ until $r=0.8$ and for $\eta=0.4$ is positive until $r=0.9$, after it turns negative due to the change in $r$ at $\lambda\geq-12.5$, but it is different when $\lambda\leq-12.6$ in fig. \ref{fig16}(b). From fig. \ref{fig16}, we observe the transverse pressure gradient $\frac{dp_t}{dr}$, which is independent of $\eta$, is negative for $\lambda\geq-12.5$ and positive for $\lambda\leq-12.6$ at $r_0=0.5$ for the specified values of $r$.

\begin{table}[htbp]
	
	\centering
	
	\begin{tabular}{|l|c|c|c|}
		\hline
		Energy    & Terms   & $\lambda =-12.5$ & $\lambda = -12.6$  \\
		conditions & & & \\
		\hline
		& $p_r$ & $\leq 0$ & $\geq 0$ \\
		
		& $p_t$ & $\geq 0$ & $\leq 0$ \\
		
		WEC     & $\rho$ & $\leq 0$ & $\geq 0$\\
		\hline
		
		NEC     & $\rho+p_r$ & $\leq 0$ & $\geq 0$\\
		& $\rho+p_t$ & $\leq 0$ & $\geq 0$\\
		
		\hline
		DEC      & $\rho-|p_r|$ & $\leq 0 $ & $\geq 0$\\
		& $\rho-|p_t|$ & $\leq 0 $ & $\geq 0$\\
		
		\hline
		SEC      & $\rho+p_r+2 p_t$ & =0 & $ =0 $\\
		& $\rho-p_r-2p_t$ & $\leq 0 $ & $\geq 0$\\
		\hline
	\end{tabular}
	\caption{Summary of a result for the energy conditions at $r_0=0.5$, $r\geq 0 ~ (0\leq r\leq 25)$, and $\eta\geq0 ~ (0\leq\eta\leq 1)$ with different value of $\lambda$ }\label{tbl1}
\end{table}

\section{Conclusions}\label{sec5}
In this research paper, we investigate an extended gravity model denoted as $f(Ric,T)$, which introduces an arbitrary connection between geometry and matter within the Finslerian framework. This connection is characterized by the trace of the stress-energy tensor. Consequently, the assumptions inherent in the $f(Ric,T)$ gravity model can lead to substantial deviations from the predictions of conventional GR and other generalized gravity models. Within this model, we derive the gravitational field equations. We proceed to construct an exponential Finslerian wormhole that is associated with the $f(Ric,T)=Ric+f(T)$ form of extended gravity theories where $f(T)=2\lambda T$. Notably the parameter $\lambda$ plays a pivotal role in the formation of this exponential Finslerian wormhole. Researchers in both the fields of physics and geometry have dedicated considerable effort to studying $f(Ric,T)$ gravity, underscoring the significance of the parameter $\lambda$ in wormhole construction.

In a recent study \cite{Singh}, researchers explored six fresh wormhole solutions within the framework of Finslerian geometry. Intriguingly, they discovered that the NEC was violated across the board for all six wormhole solutions. Additionally, they observed that both $\rho$  and $\rho+p_t$  were consistently positive for all the examined wormholes.  {F. Rahaman et al. \cite{Rahaman5} recently conducted a study where a specific mass function was assumed, leading to the derivation of an exact solution for the Finsler–Einstein field equations. The obtained solution incorporates an anisotropic matter distribution. The findings of the study indicate that all energy conditions are met satisfactorily based on the derived solution.
	Moreover, based on this work, F. Rahaman et al. \cite{Rahman} presented models of wormholes within the framework of Finslerian space-time. They discussed how these models result in the violation of energy conditions for various scenarios}. However, in our scenario, we noted a deviation from this trend: specifically for $\lambda=-12.5$ in MGT, $\rho$ and $\rho+p_t$ turned out to be negative. It's like comparing cosmic notes on the peculiarities of wormhole behavior.

Our investigation reveals that the energy conditions, including the WEC, NEC, and DEC are violated within our $f(Ric,T)$ gravity model when $\lambda\leq-12.5$. Interestingly, the SEC remains equal to zero under these conditions, as indicated in Table (\ref{tbl1}). This violation serves as compelling evidence for the wormholes' existence and exotic matter within the framework of Finslerian geometry.

Furthermore, it is noteworthy that our wormholes are found to be non-phantom as the radial EoS parameter $\omega_r$ exceeds -1.
This stands in contrast to phantom wormholes, which exhibit $\omega_r<-1$ \cite{Moraes4}. Consequently, the underlying theoretical framework is instrumental in determining the geometry, the nature of matter distribution, and the characteristics of the filled fluid.
In light of our observations, it can be concluded that our exponential Finslerian wormhole models, within the context of the $f(Ric,T)$ modified gravity theory, hold physical validity. In conclusion, the $f(Ric,T)$ gravity model provides an accurate framework for describing the properties of wormholes in the context of Finslerian space-time. These findings allow us to visualize a 3-D representation of wormhole models. Future research in this area can explore the implications of $f(Ric,T)$ gravity models on cosmic phenomena, including investigations into the impact of Finslerian wormholes on cosmological observations and their potential relevance for understanding dark energy and cosmic acceleration.


\section*{Appendix}
We constructed embedded 2-D and 3-D diagrams for the shape function equation (\ref{eq.52}) for better visualisation of the wormhole. We used an equatorial plane $\theta=\frac{\pi}{2}$ at a fixed time or $t=$constant, and $\eta=1$, from these conditions equation (\ref{eq.27})  reduce into the form
\begin{equation}\label{eq.81}
	F^2=-\left(1-\frac{b(r)}{r}\right)^{-1}dr^2-r^2d\phi^2.
\end{equation}
The above equation is represent in cylindrical coordinates as
\begin{equation}\label{eq.82}
	F^2=-dz^2-dr^2-r^2d\phi^2.
\end{equation}
In Euclidean three dimensional space $z=z(r)$ represents the embedded surface, we can rewrite equation (\ref{eq.82}) as
\begin{equation}\label{eq.83}
	F^2=-\left(1+\left(\frac{dz}{dr}\right)^2\right)dr^2-r^2d\phi^2,
\end{equation}
comaparing equation (\ref{eq.81}) and equation (\ref{eq.83})
\begin{equation}\label{eq.84}
	\frac{dz}{dr}=\pm\sqrt{\left(1-\frac{b(r)}{r}\right)^{-1}-1},
\end{equation}
using equation (\ref{eq.84}) we plotted the embedded surface of the wormhole.



\begin{thebibliography}{55}
	\bibitem{Weyl}
	Weyl H 1921 Feld und Materie {\it Annalen der Physik} \textbf{370} 541-563
	
	\bibitem{Einstein}
	Einstein A and Rosen N 1935 The particle problem in the general theory of relativity \textit{Physical review} \textbf{48} 73-77
	
	\bibitem{Fuller}
	Fuller Robert W and Wheeler John A 1962 Causality and Multiply Connected Space Time \textit{Physical Review} \textbf{128} 919-929
	
	\bibitem{Morris}
	Morris M S and Thorne K S 1988 Wormholes in space-time and their use for interstellar travel: a tool for teaching general relativity \textit{American Journal of Physics} \textbf{56} 395
	
	\bibitem{Hochberg}
	Hochberg D and Visser M 1999 General Dynamic Wormholes and Violation of the Null Energy Condition \textit{arXiv:gr-qc/9901020}
	
	\bibitem{Galloway}
	Galloway Gregory J 1995 On the topology of the domain of outer communication \textit{Classical and Quantum Gravity} \textbf{12} L99
	
	\bibitem{Friedman}
	Friedman J L, Schleich K and Witt D M 1993 Topological Censorship \textit{Physical Review Letters} \textbf{71} 1486-1489
	
	\bibitem{Lobo}
	Francisco S N Lobo and Oliveira Miguel A 2009 Wormhole geometries in $f(R)$ modified theories of gravity \textit{Physical Review D} \textbf{80} 104012
	
	\bibitem{Errehymy}
	Abdelghani E, Sunil Kumar M, Sudan H, Mohammed D, Haifa I A and Abdel-Haleem A 2023 Modeling Wormholes Generated by Dark Matter Galactic Halos in Modified Gravity \textit{Annalen der Physik } \textbf{535} 2300178
	
	\bibitem{Errehymy1}
	Abdelghani E, Sudan H, Maurya S K,  Chevarra H and Mohammed D 2023 Spherically symmetric traversable wormholes in the torsion and matter coupling gravity formalism \textit{Physics of the Dark Universe}  \textbf{41} 101258
	
	\bibitem{Harko}
	Harko T 2011 $f(R,T)$ gravity \textit{Physical Review D} \textbf{84} 024020
	
	\bibitem{Baffou}
	Baffou E H,  Houndjo M J S, Rodrigues M E,  Kpadonou A V and Tossa J 2015 Cosmological Evolution in $f(R,T)$ theory with Collisional Matter \textit{Physical Review D} \textbf{92} 084043
	
	\bibitem{Sahoo}
	Moraes P H R S and Sahoo P K 2017 Modeling wormholes in $f(R,T)$ gravity \textit{Physical Review D} \textbf{96} 044038
	
	\bibitem{Najafi}
	Najafi S, Rostami T and  Jalalzadeh S 2015 Five dimensional cosmological traversable wormhole \textit{Annals of Physics} \textbf{354} 288-297
	
	\bibitem{Paul1}
	João Luís R and Paul Martin K 2022 Non-exotic traversable wormhole solutions in linear f (R, T) gravity \textit{The European Physical Journal C} \textbf{82} 1154
	
	\bibitem{Nailya}
	João Luís R, Nailya G and Francisco S N Lobo 2023 Non-exotic traversable wormholes in $f(R,T_{ab}T^{ab})$ gravity \textit{The European Physical Journal C} \textbf{83} 1040
		
	\bibitem{Mustafa}
	Mustafa G, Maurya S K and  Hussain I 2023 Relativistic Wormholes in Extended Teleparallel Gravity with Minimal Matter Coupling \textit{Fortschritte der Physik}  \textbf{71} 2200119
	
	\bibitem{Mustafa1}
	Mustafa G, Mauryaand  S K and Saibal Ray 2023 Relativistic wormhole surrounded by dark matter halos in symmetric teleparallel gravity \textit{Fortschritte der Physik} \textbf{71} 2200129
	
	\bibitem{Moraes1}
	Moraes P H R S 2019 An exponential shape function for wormholes in modified gravity \textit{Chinese Physics Letters} \textbf{36} 120401
	
	\bibitem{Moraes2}
	Moraes P H R S and Sahoo P K 2019 Wormholes in exponential $f(R,T)$ gravity \textit{European Physical Journal C} \textbf{79} 677
	
	
	\bibitem{Banerjee}
	Banerjee A, Singh K N, Jasim M K and  Rahaman F 2020 Conformally symmetric traversable wormholes in f(R,T) gravity \textit{Annals of Physics}  \textbf{422} 168295
	
	\bibitem{Mudassar}
	Mudassar R and  Khalid S 2021 Thermodynamics of traversable wormholes in f(R,T) gravity \textit{International Journal of Geometric Methods in Modern Physics}  \textbf{18} 2150175
	
	\bibitem{Godani}
	Godani N 2022 Wormhole solutions in f(R, T) gravity \textit{New Astronomy} \textbf{94} 101774
	
	\bibitem{Bhatti}
	Bhatti M Z, Yousaf Z and Nazir M 2023 Static cylindrically symmetric wormhole models  in f(R,T) gravity \textit{ New Astronomy} \textbf{98} 101897
	
	
	\bibitem{Chang}
	Chang Z and Li X 2008 Modified Newton’s gravity in Finsler space as a possible alternative to dark matter hypothesis \textit{Physical Review B} \textbf{668} 453-456
	
	\bibitem{Voicu}
	Voicu N and Pfeifer C 2021 Finsler Gravity \textit{Modified Gravity and Cosmology} ed E N Saridakis et al (Springer) 243–259
	
	\bibitem{Claus}
	Claus Lämmerzahl and Volker Perlick 2018 Finsler geometry as a model for relativistic gravity \textit{International Journal of Geometric Methods in Modern Physics} \textbf{15} 1850166
	
	\bibitem{Sergiu}
	Vacaru S I 2012 Principles of Einstein–Finsler Gravity and Perspectives in Modern Cosmology \textit{International Journal of Modern Physics D} \textbf{21} 1250072
	
	\bibitem{Tawfik}
	Tawfik A and Dabash T F 2023 Born reciprocity and discretized Finsler structure: An approach to quantize General Relativity curvature tensors on three-sphere \textit{International Journal of Modern Physics D} \textbf{32} 2350068
	
	\bibitem{Tawfik1}
	Tawfik A and  Dabash T F 2023 Born reciprocity and relativistic generalized uncertainty principle in Finsler structure: Fundamental tensor in discretized curved space-time \textit{International Journal of Modern Physics D} \textbf{32} 2350060
	
	\bibitem{Tawfik2}
	Tawfik A and  Dabash T F 2023 Timelike geodesic congruence in the simplest solutions of general relativity with quantum-improved metric tensor \textit{International Journal of Modern Physics D} \textbf{32} 2350097
	
	\bibitem{Sumita}
	Sumita B, Shounak G, Nupur P and Farook R 2020 Study of gravastars in Finslerian geometry \textit{The European Physical Journal Plus}  \textbf{135} 185
	
	\bibitem{Krishna}
	Krishna P D and Ujjal D 2023 Possible existence of traversable wormhole in Finsler–Randers
	geometry \textit{The European Physical Journal C}  \textbf{83} 821
	
	\bibitem{Praveen}
	Narasimhamurthy S K and Praveen J 2024 Cosmological constant roll of inflation within Finsler-
	Barthel-Kropina geometry: a geometric approach to early universe dynamics \textit{New Astronomy} \textbf{108} 102187
	
	
	\bibitem{Rahman}
	Rahman F, Paul N, Banerjee A, De S S, Ray S and Usmani A A 2016 The Finslerian wormhole models \textit{European Physical Journal C} \textbf{76} 246
	
	\bibitem{Clifton}
	Clifton T, Ferreira Pedro G,  Padilla A and  Skordis C 2012 Modified gravity and cosmology \textit{Physics Reports} \textbf{513} 1-189
	
	\bibitem{Yashwanth}
	Yashwanth B R, Narasimhamurthy S K and Nekouee Z  2023 Wormhole Models for f(R,T) Gravity
	in Finsler Space-Time \textit{arXiv:gr-qc/2302.09788v2}
	
	
	\bibitem{Manjunatha}
	Manjunatha H M and  Narasimhamurthy S K 2022 The wormhole model with an exponential shape function in the Finslerian framework \textit{Chinese Journal of Physics} \textbf{77} 1561-1578
	
	\bibitem{Nekouee1}
	Nekouee Z,  Narasimhamurthy S K, Manjunatha H M  and S K Srivastava 2022 Finsler–Randers model for anisotropic constant-roll inflation \textit{The European Physical Journal Plus} \textbf{137} 1388
	
	\bibitem{Manjunatha1}
	Manjunatha H M, Narasimhamurthy S K and S K Srivastava 2023 Finslerian analogue of the Schwarzschild–de Sitter space-time \textit{Pramana} \textbf{97} 90
	
	\bibitem{Akbar-Zadeh}
	Akbar-Zadeh H 1988 Sur les espaces de Finsler a courbures sectionnelles constantes \textit{Academie Royale de Belgique Bulletin de la Classe des Sciences} \textbf{74} 281-322
	
	\bibitem{Li1}
	Li X and Chang Z 2014 Exact solution of vacuum field equation in Finsler spacetime \textit{Physical Review D} \textbf{90} 06404
	
	\bibitem{Chowdhury}
	Roy Chowdhury S, Deb D, Rahaman F, Ray S and Guha B K 2019 Anisotropic strange
	star inspired by Finsler geometry \textit{International Journal of Modern Physics D} \textbf{29} 2050001
	
     \bibitem{Wang}
     Wang H C 1947 On Finsler Spaces with Completely Integrable Equations of Killing
     \textit{Journal of the London Mathematical Society} \textbf{5} 1-22
	\bibitem{Konoplya}
	Konoplya R A 2018 How to tell the shape of a wormhole by its quasinormal modes \textit{Physics Letters B} \textbf{784} 43-49
	
	\bibitem{Mishra}
	Mishra A K, Sharma U K, Dubey V C and Pradhan A 2020 Traversable wormholes in $f(R,T)$ gravity \textit{Astrophysics and Space Science} \textbf{365} 34
	
	\bibitem{Randers}
	Randers G 1941 On an Asymmetrical Metric in the Four-Space of General Relativity \textit{Physical Review} \textbf{59} 195.
	
		\bibitem{XLi}
	Li X and Chang Z 2012 Symmetry and special relativity in Finsler space time with constant curvature \textit{Differential Geometry and its Applications} \textbf{30} 737
	
	\bibitem{Mak}
	Mak M K and Harko T 2003 Anisotropic stars in general relativity. \textit{Proceedings of the Royal Society of London A} \textbf{459} 393-408
	
		\bibitem{Singh}
	Singh K, Rahaman F, Deb D and Maurya S K 2023 Traversable Finslerian wormholes supported by phantom energy \textit{Frontiers in Physics} \textbf{10} 1038905
	
	\bibitem{Raychaudhuri}
	Stavrinos P C and  Alexiou M 2018 Raychaudhuri equation in the Finsler–Randers space-time and generalized scalar-tensor theories  \textit{International Journal of Geometric Methods in Modern Physics} \textbf{15} 1850039
	
	\bibitem{Visser}
	Visser M 1996 Lorentzian Wormholes: From Einstein to Hawking \textit{AIP Press Springer New York}
	
	
	
	\bibitem{Chanda}
	Chanda A,  Dey S and Paul B C 2021 Morris-Thorne Wormholes in $f(R,T)$ modified theory of gravity \textit{General Relativity and Gravitation} \textbf{53} 78
	
	\bibitem{Carvalho}
	Carvalho G A, Lobato R V,  Moraes P H R S,  Arbañil José D V,  Marinho Jr R M,  Otoniel E and Malheiro M 2017 Stellar equilibrium configurations of white dwarfs in the $f(R,T)$ gravity \textit{European Physical Journal C} \textbf{77} 871
	
	\bibitem{Cataldo}
	Cataldo M and Campo S 2012 Two-fluid evolving Lorentzian wormholes \textit{Physical Review D} \textbf{85} 104010
	
	
	
	\bibitem{Rahaman5}
	Rahaman F, Paul N, De S S, Ray S and Abdul Kayum Jafry M 2015 The Finslerian compact star model \textit{The European Physical Journal C} \textbf{75} 564
	
	\bibitem{Moraes4}
	Moraes P H R S, Sahoo P K,  Taori Barkha and  Sahoo Parbati 2019 Phantom  energy-dominated universe as a transient stage in $f(R)$ cosmology \textit{International Journal of Modern Physics D} \textbf{28} 1950124
	
	

	
\end{thebibliography}
\end{document}